\documentclass[12pt,a4paper]{article}
\usepackage{authblk}
\usepackage[utf8]{inputenc}
\usepackage[T1]{fontenc}
\usepackage[numbers,sort&compress]{natbib}
\usepackage{amsmath}
\usepackage{amsfonts}
\usepackage{amssymb}
\usepackage{makeidx}
\usepackage{graphicx}
\usepackage{subcaption}
\usepackage{float}
\usepackage{appendix}
\usepackage[colorlinks=true,linktocpage=true,
linkcolor=blue,citecolor=blue]{hyperref}
\title{\textbf{Flow of charge and heat in thermal QCD within the weak magnetic field limit: A BGK model approach}}
\author[1]{\textbf{Anowar Shaikh}\footnote{ \href{mailto:anowar.19dr0016@ap.iitism.ac.in}{anowar.19dr0016@ap.iitism.ac.in}}}
\author[2]{\textbf{Shubhalaxmi Rath}\footnote{ \href{mailto:shubhalaxmi@iitb.ac.in}{shubhalaxmi@iitb.ac.in}}}
\author[2]{\textbf{Sadhana Dash}\footnote{ \href{mailto:sadhana@phy.iitb.ac.in}{sadhana@phy.iitb.ac.in}}}
\author[1]{\textbf{Binata Panda}\footnote{ \href{mailto:binata@iitism.ac.in}{binata@iitism.ac.in}}}
\affil[1]{ Department of  Physics\\ Indian Institute of Technology (Indian School of Mines) Dhanbad, Jharkhand 826004, India}
\affil[2]{ Department of  Physics\\ Indian Institute of Technology Bombay, Mumbai 400076, India}
\date{}
\begin{document}
	\maketitle
	\begin{abstract}
We have computed the charge and heat transport coefficients of hot QCD matter by solving the relativistic Boltzmann transport equation using the BGK model approximation with a modified collision integral in the weak magnetic field regime. This modified collision integral enhances both charge and heat transport phenomena which can be understood by the large values of the above-mentioned coefficients in comparison to the relaxation collision integral. We have also presented a comparative study of coefficients like the electrical conductivity ($\sigma_{el}$), Hall conductivity ($\sigma_{H}$), thermal conductivity ($\kappa$) and Hall-type thermal conductivity($\kappa_{H}$) in weak and strong magnetic fields in the  BGK model approximation. The effects of weak magnetic field and finite chemical potential on the transport coefficients have been explored using a quasiparticle model. Moreover, we have also studied the effects of weak magnetic field and finite chemical potential on Lorenz number, Knudsen number, specific heat, elliptic flow and Wiedemann-Franz law. 
\end{abstract}

\section{Introduction}
The possible existence of the de-confined state of quarks and gluons, known as Quark Gluon Plasma (QGP)  has been theoretically predicted by quantum chromodynamics (QCD). The creation of such a state has been achieved in the  heavy ion collision experiments at the Relativistic Heavy Ion Collider (RHIC) at BNL and the Large Hadron Collider (LHC) at CERN. 
In such collisions, strong magnetic fields are expected to be produced in the perpendicular direction to the collision plane. The strength of the magnetic field can be expressed in terms of the pion mass scale as $eB=m^2_{\pi}$ at RHIC \cite{skokov2009estimate} and  $eB=15m^2_{\pi}$ at LHC \cite{bzdak2012event}. The creation of QGP in laboratory is of great phenomenological significance. The extensive understanding of various aspects of QGP in the presence of magnetic field is becoming increasingly relevant. In recent years, various properties of QGP in the presence of magnetic field have been studied by different research groups. Some of them are chiral magnetic effect \cite{fukushima2008chiral, kharzeev2008effects}, axial magnetic effect \cite{braguta2014temperature, chernodub2014condensed}, magnetic and inverse magnetic catalysis \cite{gusynin1994catalysis,lee1997chiral}, the nonlinear electromagnetic current \cite{kharzeev2014chiral,satow2014nonlinear}, chiral vortical effect \cite{kharzeev2011testing}, the axial Hall current \cite{pu2015chiral}, refractive indices and decay constant \cite{fayazbakhsh2013weak,fayazbakhsh2012properties}, dispersion relation in magnetized thermal QED \cite{sadooghi2015magnetized}, photon and dilepton productions from QGP \cite{van2011thermal,shen2014thermal,tuchin2013magnetic}, thermodynamic and magnetic properties \cite{rath2017one,bandyopadhyay2019pressure,rath2019thermomagnetic,karmakar2019anisotropic}, heavy quark diffusion \cite{fukushima2016heavy}, and magnetohydrodynamics \cite{roy2015analytic,inghirami2016numerical}. 

The transport coefficients like, electrical conductivity ($\sigma_{el}$) and thermal conductivity ($\kappa$) carry the information about the charge and heat transport in the medium, respectively. Electrical conductivity plays an important role for the study of chiral magnetic effect \cite{fukushima2008chiral, kharzeev2008effects}, emission rate of soft photon \cite{kapusta2006finite} etc. The effect of magnetic field on electrical conductivity has already been studied via different approaches, such as the dilute instanton-liquid model \cite{nam2012electrical}, the diagrammatic method using the real-time formalism \cite{hattori2016electrical}, the quenched SU$(2)$ lattice gauge theory \cite{buividovich2010magnetic}, the effective fugacity approach \cite{kurian2017effective} etc. The studies on $\kappa$ of a hot QCD matter in a strong magnetic field have been done in \cite{kurian2019transport,rath2019revisit}. In literature, different approaches are available for computing these coefficients, such as the relativistic Boltzmann transport equation \cite{muronga2007relativistic,puglisi2014electric,thakur2017shear,yasui2017transport}, the Chapman-Enskog approximation \cite{Mitra2016,Mitra2017}, the correlator technique using Green-Kubo formula \cite{nam2012electrical,Greif2014,Feng2017}, the quasiparticle model \cite{Denicol2018} etc. 

Recently, in a work \cite{Rath2021}, the effects of weak magnetic field and finite chemical potential on the transport of charge and heat in a hot QCD matter have been studied by calculating the transport coefficients, such as the electrical conductivity ($\sigma_{el}$), the Hall conductivity ($\sigma_{H}$), the thermal conductivity ($\kappa$) and the Hall-type thermal conductivity ($\kappa_{H}$). The transport coefficients have been calculated by solving the relativistic Boltzmann transport equation in weak field regime, and the complexities of the collision term were avoided by considering a mean free-path treatment in the kinetic theory approach within the relaxation time approximation (RTA). 
However, in such models, the charge is not conserved instantaneously, but only on an average basis over a cycle. This can be avoided by considering a modified collision term as introduced in Bhatnagar-Gross-Krook (BGK) model \cite{Bhatnagar1954}. The effects of this modified collision integral and strong magnetic field on charge and heat transport for hot QCD medium were recently  explored in \cite{Khan2021}. In BGK collisional kernel, the particle number is conserved instantaneously, and has been studied on plasma instabilities in \cite{Schenke2006}. Moreover, one can consider QGP as a system of massive non-interacting quasiparticles in the quasiparticle model framework \cite{goloviznin1993refractive, Peshier1996}. For the study of QGP, this type of model is suitable as the medium formed in heavy ion collision behaves like a strongly coupled system. Later different research groups consider this model for different scenarios such as Namb-Jona-Lasinio model \cite{Fukushima2004,Ghosh2006,Abuki2009}, thermodynamically consistent quasiparticle model \cite{Bannur2007,Bannur2007a}, quasiparticle model in a strong magnetic field \cite{Rath2020,Rath2021a} and quasiparticle model with Gribov-Zwanziger quantization \cite{Su2015,Florkowski2016}. 

In this paper, we plan to explore the effects of weak magnetic field and finite chemical potential for a hot and dense QCD matter by estimating their respective response functions, {\em viz.} $\sigma_{el}$, $\sigma_{H}$, $\kappa$ and $\kappa_{H}$. A kinetic theory approach with the BGK collision term in the relativistic Boltzmann transport equation has been used to calculate the above coefficients followed by the estimation of the  Lorenz number, Knudsen number, specific heat, elliptic flow coefficient etc.  We also showed how these transport coefficients change as we shift to the quasiparticle description of partons, where the rest masses are replaced by the masses generated in the medium.  A comparative discussion of the  results with the previous two works \cite{Rath2021, Khan2021}  is also presented.

The rest of the paper has been organized as follows. Section $2$ is dedicated to the calculation of charge and heat transport coefficients from the relativistic Boltzmann transport equation with a BGK collision integral in the weak magnetic field regime. In section $3$, we have discussed the results of these different kinds of conductivity, considering the rest mass of the quarks. The quasiparticle model is discussed in weak magnetic field and finite chemical potential in section $4.1$. The results of the above-stated transport coefficients are discussed in section $4.2$. Section $5$ describes various applications and their results obtained from these coefficients, such as Lorenz number, Knudsen number, specific heat, elliptic flow coefficient etc. Finally, section $6$ summarizes the results. 

\noindent \textbf{Notations and conventions}: Here the covariant derivative $\partial_\mu$ and $\partial^{(p)} _\mu$ are written for $\frac{\partial}{\partial x^\mu}$ and 
$\frac{\partial}{\partial p^\mu}$ respectively. The fluid four velocity  $u^\mu=(1,0,0,0)$ is normalized to unity in the rest frame  $(u^\mu u_\mu=1)$. Throughout this paper, the subscript $f$ stands for the flavour index with $f$=u, d, s. Moreover, $q_f$, $g_f$, and $ \delta f_{f}$ ($\delta \bar {f}_{f}$) are the electric charge, degeneracy factor and the infinitesimal change in the distribution function for the quark (antiquark) of the $f^{th}$ flavour, respectively. In equation (\ref{B}), $\sigma_0$, $ \sigma_1 $ and $ \sigma_2 $ are the various components of the electrical conductivity tensor and $\mathbf{b}=\textbf{B}/B$ indicates the direction of magnetic field. In equation (\ref{C}), $\epsilon^{ij}$ denotes an antisymmetric $2\times 2$ unit matrix. The Lorentz force is defined as $\mathbf{F}=q(\mathbf{E}+\mathbf{v}\times \mathbf{B})$ and the components of ${F}^{\mu \nu}$ are related to electric and magnetic fields as ${F}^{0i}=E^i$, $F^{i0}=-E^i$ and $F^{ij}=\frac{1}{2}\epsilon^{ijk}B_k$. Here $m_f$ denotes the current quark masses (which are 3 MeV, 5 MeV and 100 MeV for up, down and strange quarks, respectively). In this work, $g_g=2(N^2_c-1)$ represents the gluonic degrees of freedom, and $g_f(g_{\bar{f}})=2N_c N_f$ denotes the quark (antiquark) degrees of freedom, where $N_{f}=3$ and $N_c=3$ are the number of flavours and colors. 

\section{Charge and heat transport properties of a QCD medium in the presence of magnetic field: A BGK model approach}
 The Boltzmann transport equation for a single particle distribution function is given by	
 \begin{equation}\label{65} 	
 \frac{\partial f_f}{\partial t}+ \frac{\bf{p}}{m}.\nabla f_f + \bf{F}.\frac{\partial \textit{f}_\textit{f}}{\partial \bf{p}}=\left(\frac{{\partial \textit{f}_\textit{f}}}{\partial \textit{t}}\right)_c =C\bigl[\textit{f}_\textit{f}\hspace{1mm}\bigr], 
 \end{equation}
where \textbf{F}  is the force field acting on the particles in the medium and m is the particle's mass. The right-hand side term arises due to collisions between particles in the medium. The zero value of this term indicates a collisionless system referred to as the Vlasov equation. Even for the simplest solution for the above equation (\ref{65}), one faces various difficulties bacause of the complicated nature of the collision term. The term $\left(\frac{\partial f_\textit{f}}{\partial t}\right)_c$ represents the instantaneous change in the distribution function due to collisions between particles. In order to obtain a complete solution, we will start by discussing some mathematical models by which we can adequately treat the collision term. 

In the case of relaxation time approximation (RTA), where the collision term is being replaced by a relaxation term having a form as given below
 \begin{equation}	
 	\left( \frac{\partial f_\textit{f}}{\partial t}\right) _c=-\frac{1}{\tau(p)}\left( f_\textit{f}(x,p,t)-f_{eq,f}(p)\right),  	
 \end{equation} 
where we can write $f_\textit{f}(x,p,t)=f_{eq,f}(p)+\delta f_f(x,p,t)$ and $\tau (p)$ 
signifies the relaxation time for occurring collisions which forge ahead the 
distribution function to equilibrium state. In order to allow linearization of the Boltzmann equation, we can assume the quark distribution function is close to equilibrium with a small deviation from equilibrium. Here the term $\frac{1}{\tau }$ acts as a damping frequency. The instantaneous conservation of charge appears to be an elementary issue of this model. To shed the crunch, Bhatnagar-Gross-Krook (BGK) model introduces a new type of collision kernel, which is given as
 \begin{equation}
 	\left( \frac{\partial f_\textit{f}}{\partial t}\right)_c=-\frac{1}{\tau(p)}\left( f_f(x,p,t)-\frac {n(x,t)}{n_{eq}}f_{eq,f}(p)\right), 	
 \end{equation}
 where $n(x,t)$ is known as fluctuating density or perturbed density calculated as $\int f_f(x,p,t) d^3p $ which gives zero value after integrating over momenta, thus indicating instantaneous conservation of particle number during collisions in the system. Here we are considering the general relativistic covariant form of the Boltzmann transport equation for a medium of quarks and gluons given as
 \begin{equation}\label{D}
 	p^\mu \partial_\mu f_f+qF^{\mu \nu} p_\nu \partial^{(p)} _\mu f_f=C[f_\textit{f}]=-  \frac{p^\mu u_\mu}{\tau(p)}\left( f_\textit{f}(x,p,t)-\frac {n(x,t)}{n_{eq}}f_{eq,f}(p)\right),  
 \end{equation}
 where $F^{\mu \nu}$ is the electromagnetic field strength tensor representing the external electric and magnetic fields applied to the system. We will now see in forthcoming sections how the above-mentioned
 collision integral affects the solution of the relativistic Boltzmann transport equation in the presence of a weak magnetic field. We also discuss the consequences on the transport of
 current and heat in terms of their respective transport
 coefficients, such as $\sigma_{el}$, $\sigma_{H}$, $\kappa$ and $\kappa_{H}$ 
 and the derived coefficients from them, namely Lorenz number,
 Knudsen number, elliptic flow and specific heat. The next two sections of the present manuscript are arranged as follows.  We will begin with the formalism of electric charge transport and calculation of corresponding transport coefficients ($\sigma_{el}$, $\sigma_{H}$) within the weak magnetic field regime in section 2.1. Further, section 2.2 contains the formalism of heat transport phenomena and the calculation of associated transport coefficients ($\kappa$, $\kappa_{H}$) within the weak magnetic field regime. 

\subsection{Charge transport properties (electrical conductivity and Hall conductivity)}
In an attempt to see the effects of the magnetic field on the charge transport phenomena, we need to calculate the electric current density of the magnetized medium. When a thermal QCD medium containing quarks, antiquarks and gluons of different flavours gets in contact with an external electric field, the medium gets infinitesimally disturbed. Due to this electric field, there is an induced current density in the medium, whose spatial component (\textbf{${J^{i}}$}) is directly proportional to the electric field (${E}$) with a proportionality constant, known as electrical conductivity ($\sigma_{el}$) and is given by
 \begin{equation}\label{A}
 	J^{i}=\sum_{f}  g_{f}q_{f}\int \frac {d^3p}{(2\pi)^3} \frac{p^i}{\omega_f}(\delta f_f(x,p)+\delta \bar{f}_f(x,p)),
 \end{equation}
where $J^{i}$ represents the spatial part of the current density vector. At first, we have to calculate the nonequilibrium part of the distribution function from equation (\ref{D}) to get the current density. In the presence of an electromagnetic field, the general form of spatial current density can be written as
 \begin{equation}\label{B}
 	J^i=\sigma^{ij}E_j=\sigma_0\delta^{ij}E_j+\sigma_1\epsilon^{ijk}b_kE_j+\sigma_2b^ib^jE_j.
 \end{equation}
 The equation $ (\ref{B}) $ can be rewritten in a case where the electric field and magnetic field are both perpendicular to each other as 
 \begin{equation}\label{C}
 	J^i=\sigma^{ij}E_j=(\sigma_{el}\delta^{ij}+\sigma_H\epsilon^{ij})E_j.
 \end{equation}
 Here $\sigma_0$ and $\sigma_1$ denote electrical conductivity ($\sigma_{el}$)
 and Hall conductivity ($\sigma_H$), respectively. Now one can obtain the expressions for electrical conductivity and Hall conductivity 
 with the help of equations (\ref{A}) and (\ref{C}). 
 In order to do that, we start with the relativistic Boltzmann transport equation (RBTE), {\em i.e.} 
 eq. (\ref{D}), which can be rewritten as
 \begin{equation}\label{E}
 	p^\mu\frac{\partial f_f}{\partial x^\mu}+ qF^{\mu\nu}p_\nu \frac{\partial f_f}{\partial p^\mu}=C[f_f]=-\frac{p_\mu u^\mu}{\tau_f}\left( f_f-n_f n^{-1}_{eq,f}f_{eq,f}\right),	
 \end{equation}
 where  $\tau_f$ is the relaxation time, {\em i.e.} the time required to bring the perturbed system back to its equilibrium state and $\nu_f=\frac{1}{\tau_f}$ indicates the collision frequency of the medium. The equilibrium distribution functions for $f^{th}$ flavour of quark and antiquark are given as
 \begin{equation}
 	f_{eq,f}=\frac{1}{e^{\beta(\omega_f-\mu_f)}+1},	
 \end{equation}
 \begin{equation}
 	\bar{f}_{eq,f}=\frac{1}{e^{\beta(\omega_f+\mu_f)}+1},
 \end{equation}
 respectively, where $\omega_f=\sqrt{(\textbf{p}^2+m^2_f)}$, $\beta=\frac{1}{T}$, and $\mu_f$ represents the chemical potential of $f^{th}$ flavour of quark. In a weak magnetic field regime, the temperature scale is more dominant than the magnetic field scale in equilibrium. For this reason, we neglect the effects of Landau quantization on phase space and scattering processes. The expression for $\tau_f$ can be calculated \cite{hosoya1985transport} as
 \begin{equation}
 	\tau_{f}(T)=\frac{1}{5.1T\alpha^2_s \log(1/\alpha_s)\left[1+0.12(2N_f+1)\right] }.
 \end{equation}
 Here $\alpha_s$ is the  QCD running coupling constant, which is a function of both magnetic field and chemical potential and it has the following \cite{ayala2018thermomagnetic} form, 
 \begin{equation}
 	\alpha_s(\Lambda^2,eB)=\frac{\alpha_s(\Lambda^2)}{\left[ 1+b_1 \alpha_s(\Lambda^2)\ln(\frac{\Lambda^2}{\Lambda^2+eB})\right] },
 \end{equation}
 where
 \begin{equation}
 	\alpha_s(\Lambda^2)=\frac{1}{b_1 \ln\left(\frac{\Lambda^2}{\Lambda^2_{\overline{MS}}}\right) },
 \end{equation} 
 with $b_1=\frac{(11N_c-2N_f)}{12\pi}$, $\Lambda_{\overline{MS}}=0.176$ GeV and $\Lambda=2\pi\sqrt{T^2+\left(\frac{\mu_f^2}{\pi^2}\right)}$. In the collision term, $n_f$ and $n_{eq,f}$ signify the perturbed density and equilibrium density, respectively for the $f^{th}$ flavour with a degeneracy of $g_f$. Both can be calculated from the equations given below, 
 \begin{equation}
 	n_f=g_f\int \frac{d^3p}{(2\pi)^3}(f_{eq,f}+\delta f_f),
 \end{equation}
 \begin{equation}
 	n_{eq,f}=g_f\int \frac{d^3p}{(2\pi)^3}f_{eq,f}.
 \end{equation}
 \\The RBTE (\ref{E}) can be rewritten as
 \begin{equation}
 	\frac{\partial f_f}{\partial t}+\textbf{v}.\frac{\partial f_f}{\partial \textbf{r}}+\frac{\textbf{p.F}}{p^0}\frac{\partial f_f}{\partial p^0}+\textbf{F}.\frac{\partial f_f}{\partial \textbf{p}}=C[f_f]=-\frac{p_\mu u^\mu}{\tau_f}(f_f-n_f n^{-1}_{eq,f}f_{eq,f}).
 \end{equation}
 Now we consider the distribution function to be spatially homogeneous and also in a steady state condition, which gives us the freedom to neglect the first two terms by taking $\frac{\partial f_f}{\partial \textbf{r}}=0$, $\frac{\partial f_f}{\partial t}=0$ and finally the above equation is reduced to
 \begin{equation}
 	\textbf{v.F}\frac{\partial f_f}{\partial p^0}+\textbf{F}.\frac{\partial f_f}{\partial \textbf{p}}=C[f_f]=-\frac{p_\mu u^\mu}{\tau_f}(f_f-n_f n^{-1}_{eq,f}f_{eq,f}).
 \end{equation}
 By taking electric and magnetic fields in a perpendicular direction to each other {\em i.e.} electric field along x-direction ($\textbf{E}=E\hat x$) and magnetic field along z-direction ($\textbf{B}=B\hat z$), the above equation appears to be
 \begin{equation}\label{H}
 	\tau_{f}qEv_x\frac{\partial f_f}{\partial p^0}+\tau_{f}qBv_y\frac{\partial f_f}{\partial p_x}-\tau_{f}qBv_x\frac{\partial f_f}{\partial p_y}=(n_f n^{-1}_{eq,f}f_{eq,f}-f_f).	
 \end{equation}
 For a satisfactory solution to the above equation, we propose the following ansatz for the distribution function, which was first introduced in
  \cite{Feng2017} and was later followed by various research groups \cite{das2020electrical, das2019electrical,bandyopadhyay2020anisotropic,thakur2019electrical}. 
 \begin{equation}\label{F}
 	f_f=n_f n^{-1}_{eq,f}f_{eq,f}-\tau_fq\textbf{E}.\frac{\partial f_{eq,f}}{\partial \textbf{p}}-\boldsymbol{\Gamma}. \frac{\partial f_{eq,f}}{\partial \textbf{p}},
 \end{equation}
 Here $\boldsymbol{\Gamma}$ is an unknown quantity which is constructed in such a way that it depends on both electric field and magnetic field in weak magnetic field limit (it will be more evident in the forthcoming sections where we will obtain its expression). Neglecting higher-order terms of $eB$ in equation (\ref{F}), one can easily show that the unknown quantity $\boldsymbol{\Gamma}$ should be a function of $eB$. 
 
Now with the help of an assumption that the quark distribution function is in the neighbourhood of equilibrium, we can evaluate these quantities as follows, 
 \begin{equation}\label{G}	
 	\left.	\begin{aligned}
 		\frac{\partial f_{eq,f}}{\partial p_x}=-\beta v_x f_{eq,f}(1-f_{eq,f}) \\
 		\frac{\partial f_{eq,f}}{\partial p_y}=-\beta v_y f_{eq,f}(1-f_{eq,f}) \\
 		\frac{\partial f_{eq,f}}{\partial p_z}=-\beta v_z f_{eq,f}(1-f_{eq,f})
 	\end{aligned}
 	\right\}.
 \end{equation}
 With the help of equations (\ref{F}) and (\ref{G}) at high temperature, eq. (\ref{H}) can be simplified as
 \begin{equation}
 	\tau_f qEv_x\frac{\partial f_f}{\partial p^0}+ \beta f_{eq,f}(\Gamma_x v_x+\Gamma_y v_y+\Gamma_z v_z)- qB\tau_f(v_x\frac{\partial f_f}{\partial p_y}-v_y\frac{\partial f_f}{\partial p_x})=0.
 \end{equation}
 Solving the above equation, we have\footnote{Refer appendix A of \cite{Rath2021} for detailed mathematical calculations.}
 \begin{equation}\label{x3}
 	\left.	\begin{aligned}
 		\Gamma_x=\frac{qE\tau_f(1-\omega^2_c \tau^2_f)}{(1+\omega^2_c \tau^2_f)} \\
 		\Gamma_y=-\frac{2qE\omega_c \tau^2_f}{(1+\omega^2_c \tau^2_f)} \\
 		\Gamma_z=0
 	\end{aligned}
 	\right\}.
 \end{equation}
 Putting the values of  $\Gamma_x$, $\Gamma_y$ and $\Gamma_z$ in the ansatz, we get 
 \begin{multline}\label{I}
 	(f_f-n_f n^{-1}_{eq,f}f_{eq,f})= 2qE\beta v_x\left( \frac{\tau_f}{(1+\omega^2_c \tau^2_f)}\right) f_{eq,f}(1-f_{eq,f}) \\ - 2qE\beta v_y\left (\frac{\omega_c \tau^2_f}{(1+\omega^2_c \tau^2_f)}\right)
 	f_{eq,f}(1-f_{eq,f}). 
 \end{multline}
 One can reduce the LHS of eq. (\ref{I}) to \footnote{Refer appendix A of \cite{Khan2021} for detailed mathematical calculations.} 
 \begin{equation}\label{xxv}
 	(f_f-n_f n^{-1}_{eq,f}f_{eq,f})=\left[ \delta f_f-g_f n^{-1}_{eq,f} f_{eq,f}\int_{p} \delta f_f\right]\footnote{Here we are using the symbol of momentum integration, $\int_{p}=\int {d^3p}/{(2\pi)^3}$.} .	
 \end{equation}
 Substituting the result of eq. (\ref{xxv}) in eq. (\ref{I}) we have
 \begin{multline}\label{x1}
 	\delta f_f-g_f n^{-1}_{eq,f} f_{eq,f}\int_{p} \delta f_f=2qE\beta v_x\left( \frac{\tau_f}{(1+\omega^2_c \tau^2_f)}\right) f_{eq,f}(1-f_{eq,f}) \\ - 2qE\beta v_y\left (\frac{\omega_c \tau^2_f}{(1+\omega^2_c \tau^2_f)}\right)
 	f_{eq,f}(1-f_{eq,f}).
 \end{multline}
Neglecting higher order terms $\left(\mathcal O(\delta f_f)^2\right)$, we get the solution of this equation as
 \begin{equation}
 	\delta f_f=\left[ \delta f^{(0)}_f +g_f n^{-1}_{eq,f} f_{eq,f}\int_{ p\prime } \delta f^{(0)}_f\right] ,
 \end{equation}
 where \\
 \begin{multline}
 	\delta f^{(0)}_f=\Biggl[2qE\beta v_x\left( \frac{\tau_f}{(1+\omega^2_c \tau^2_f)}\right) f_{eq,f}(1-f_{eq,f})
 	- 2qE\beta v_y\left (\frac{\omega_c \tau^2_f}{(1+\omega^2_c \tau^2_f)}\right) \\ \times f_{eq,f}(1-f_{eq,f})\Biggr].
 \end{multline}\\
 The infinitesimal change in the quark distribution function ($\delta f_f$) (see appendix \ref{appendix A}) is obtained as
 \begin{multline}\label{x2}
 	\delta f_f=\Biggl[ 2qE\beta v_x\Biggl( \frac{\tau_f}{(1+\omega^2_c \tau^2_f)}\Biggr) f_{eq,f}(1-f_{eq,f})
 	- 2qE\beta v_y\left (\frac{\omega_c \tau^2_f}{(1+\omega^2_c \tau^2_f)}\right) \\ \times f_{eq,f}(1-f_{eq,f}) \Biggr]
 	+ g_f n^{-1}_{eq,f} f_{eq,f} \int_{p \prime} \Biggr[ 2qE\beta v_x\Biggr( \frac{\tau_f}{(1+\omega^2_c \tau^2_f)}\Biggl) f_{eq,f}(1-f_{eq,f})\\
 	- 2qE\beta v_y \Biggr (\frac{\omega_c \tau^2_f}{(1+\omega^2_c \tau^2_f)}\Biggl) f_{eq,f}(1-f_{eq,f}) \Biggl]. 
 \end{multline}
 Similarly, for antiquarks, we get 
 \begin{multline}
 	\delta {\bar{f}}_f=\Biggl[2\bar{q}E\beta v_x\Biggl( \frac{ \tau_{\bar{f}}}{(1+\omega^2_c \tau^2_{\bar{f}})}\Biggr) {\bar{f}}_{eq,f}(1-{\bar{f}}_{eq,f})
 	- 2\bar{q}E\beta v_y\Biggl (\frac{\omega_c\tau^2_{\bar{f}}}{(1+\omega^2_c \tau^2_{\bar{f}})}\Biggr) \\ \times {\bar{f}}_{eq,f}(1-{\bar{f}}_{eq,f}) \Biggr]
 	+ g_f n^{-1}_{eq,f} f_{eq,f} \int_{p \prime} \Biggl[2\bar{q}E\beta v_x\Biggl( \frac{\tau_{\bar{f}}}{(1+\omega^2_c \tau^2_{\bar{f}})}\Biggr) {\bar{f}}_{eq,f}(1-{\bar{f}}_{eq,f})\\
 	- 2\bar{q}E\beta v_y\Biggl (\frac{\omega_c\tau^2_{\bar{f}}}{(1+\omega^2_c \tau^2_{\bar{f}})}\Biggr) {\bar{f}}_{eq,f}(1-{\bar{f}}_{eq,f}) \Biggr].
 \end{multline}
 Now, replacing the values of $\delta f_f$ and $\delta {\bar{f}}_f$ in equation (\ref{A}) and comparing it with equation (\ref{C}), we get the final expressions for electrical conductivity and Hall conductivity under BGK model in weak magnetic field limit as
 \begin{multline}\label{x8}	
 	\sigma^{BGK}_{el}=\frac{\beta}{3\pi^2}\sum_{f} g_f q^2_f\int dp\frac{p^4}{\omega^2_f}\Biggl[\Biggl( \frac{\tau_f}{(1+\omega^2_c \tau^2_f)}\Biggr) f_{eq,f}(1-f_{eq,f})
 	+\Biggl (\frac{\tau_{\bar{f}}}{(1+\omega^2_c \tau^2_{\bar{f}})}\Biggr) \\ \times {\bar{f}}_{eq,f}(1-{\bar{f}}_{eq,f}) \Biggr] +   
 	\frac{2\beta}{\sqrt{3}}\sum_{f} g^2_f q^2_f n^{-1}_{eq,f}\int_{p} \frac{p}{\omega_f} f_{eq,f} \times \int_{p \prime}\frac{p \prime}{\omega_f}\Biggl[\Biggl( \frac{\tau_f}{(1+\omega^2_c \tau^2_f)}\Biggr) f_{eq,f}(1-f_{eq,f})\\
 	+\Biggl (\frac{\tau_{\bar{f}}}{(1+\omega^2_c \tau^2_{\bar{f}})}\Biggr) {\bar{f}}_{eq,f}(1-{\bar{f}}_{eq,f}) \Biggr],	
 \end{multline}
 \begin{multline}\label{x9}	
 	\sigma^{BGK}_{H}=\frac{\beta}{3\pi^2}\sum_{f} g_f q^2_f\int dp\frac{p^4}{\omega^2_f}\Biggl[\Biggl( \frac{\omega_c\tau^2_f}{(1+\omega^2_c \tau^2_f)}\Biggr) f_{eq,f}(1-f_{eq,f})
 	+\Biggl (\frac{\omega_c\tau^2_{\bar{f}}}{(1+\omega^2_c \tau^2_{\bar{f}})}\Biggr) \\ \times {\bar{f}}_{eq,f}(1-{\bar{f}}_{eq,f}) \Biggr] +
 	\frac{2\beta}{\sqrt{3}}\sum_{f} g^2_f q^2_f n^{-1}_{eq,f}\int_{p} \frac{p}{\omega_f} f_{eq,f} \times \int_{p \prime}\frac{p \prime}{\omega_f}\Biggl[\Biggl( \frac{\omega_c\tau^2_f}{(1+\omega^2_c \tau^2_f)}\Biggr) f_{eq,f}(1-f_{eq,f})\\
 	+\Biggl (\frac{\omega_c\tau^2_{\bar{f}}}{(1+\omega^2_c \tau^2_{\bar{f}})}\Biggr) {\bar{f}}_{eq,f}(1-{\bar{f}}_{eq,f}) \Biggr].
 \end{multline}
Expressions of $\sigma^{BGK}_{el}$ (eq. \ref{x8}) and $\sigma^{BGK}_{H}$ (eq. \ref{x9}) can be rewritten as
 \begin{equation}
 	\sigma^{BGK}_{el}=\sigma^{RTA}_{el}+\sigma^{Corr}_{el},
 \end{equation}
 \begin{equation}
 	\sigma^{BGK}_{H}=\sigma^{RTA}_{H}+\sigma^{Corr}_{H},
 \end{equation}
 where 
 \begin{multline}
 	\sigma^{RTA}_{el}=\frac{\beta}{3\pi^2}\sum_{f} g_f q^2_f\int dp\frac{p^4}{\omega^2_f}\Biggr[\left( \frac{\tau_f}{(1+\omega^2_c \tau^2_f)}\right) f_{eq,f}(1-f_{eq,f})\\
 	+\left (\frac{\tau_{\bar{f}}}{(1+\omega^2_c \tau^2_{\bar{f}})}\right) {\bar{f}}_{eq,f}(1-{\bar{f}}_{eq,f}) \Biggl],
 \end{multline}
 \begin{multline}
 	\sigma^{Corr}_{el}=\frac{2\beta}{\sqrt{3}}\sum_{f} g^2_f q^2_f n^{-1}_{eq,f}\int_{p} \frac{p}{\omega_f} f_{eq,f} \times \int_{p \prime}\frac{p \prime}{\omega_f}\Biggl[\left( \frac{\tau_f}{(1+\omega^2_c \tau^2_f)}\right) f_{eq,f}(1-f_{eq,f})\\
 	+\left (\frac{\tau_{\bar{f}}}{(1+\omega^2_c \tau^2_{\bar{f}})}\right) {\bar{f}}_{eq,f}(1-{\bar{f}}_{eq,f}) \Biggr].
 \end{multline}
 Similarly,\\
 \begin{multline}
 	\sigma^{RTA}_{H}=\frac{\beta}{3\pi^2}\sum_{f} g_f q^2_f\int dp\frac{p^4}{\omega^2_f}\Biggl[\left( \frac{\omega_c\tau^2_f}{(1+\omega^2_c \tau^2_f)}\right) f_{eq,f}(1-f_{eq,f})\\
 	+\left (\frac{\omega_c\tau^2_{\bar{f}}}{(1+\omega^2_c \tau^2_{\bar{f}})}\right) {\bar{f}}_{eq,f}(1-{\bar{f}}_{eq,f}) \Biggr],
 \end{multline}
 \begin{multline}	
 	\sigma^{Corr}_{H}=	\frac{2\beta}{\sqrt{3}}\sum_{f} g^2_f q^2_f n^{-1}_{eq,f}\int_{p} \frac{p}{\omega_f} f_{eq,f} \times \int_{p \prime}\frac{p \prime}{\omega_f}\Biggr[\left( \frac{\omega_c\tau^2_f}{(1+\omega^2_c \tau^2_f)}\right) f_{eq,f}(1-f_{eq,f})\\
 	+\left (\frac{\omega_c\tau^2_{\bar{f}}}{(1+\omega^2_c \tau^2_{\bar{f}})}\right) {\bar{f}}_{eq,f}(1-{\bar{f}}_{eq,f}) \Biggl].
 \end{multline}
 
\subsection{Heat transport properties (thermal conductivity and Hall-type thermal conductivity)}
Heat flow in a system is defined as the difference between energy flow and enthalpy flow. It is directly proportional to the temperature gradient through a proportionality constant known as thermal 
conductivity. To study the heat transport in a medium, we need to calculate the thermal conductivity first.
 In four-vector notation, the heat flow is given by
 \begin{equation}
 	Q_{\mu}=\Delta_{\mu \alpha}T^{\alpha\beta}u_\beta-h\Delta_{\mu \alpha}N^\alpha.
 \end{equation}
 In the above equation, the projection operator ($\Delta_{\mu \alpha}$) and the enthalpy per 
 particle ($h$) are respectively defined as
 \begin{equation}
 	\Delta_{\mu \alpha}=g_{\mu \alpha}-u_{\mu} u_{\alpha},
 \end{equation}
 \begin{equation}
 	h=(\epsilon+P)/n,
 \end{equation}   	
 where $\epsilon$, $P$ and $n$ represent the energy density, pressure and particle number density, respectively, which are given by the following equations, 
 \begin{equation}\label{P}
 	\epsilon=u_{\alpha}T^{\alpha \beta}u_{\beta},
 \end{equation}
 \begin{equation}
 	P=-\frac{\Delta_{\alpha\beta} T^{\alpha\beta}}{3},
 \end{equation}
 \begin{equation}
 	n=N^{\alpha}u_{\alpha}.
 \end{equation}
 Here, the particle flow four vector $N^{\alpha}$	and the energy-momentum tensor $T^{\alpha \beta}$ are defined as
 \begin{equation}
 	N^{\alpha}=\sum_{f} g_f \int \frac{d^3p}{(2\pi)^3}\frac{p^{\alpha}}{\omega_f}[f_f(x,p)+\bar{f}_f(x,p)],	
 \end{equation}
 \begin{equation}
 	T^{\alpha \beta}=\sum_{f} g_f \int \frac{d^3p}{(2\pi)^3} \frac{p^{\alpha} p^{\beta}}{\omega_f}[f_f(x,p)+\bar{f}_f(x,p)]. 	
 \end{equation}
 The spatial part of the heat flow four-vector is given by
 \begin{equation}\label{K}
 	Q^{i}=\sum_{f} g_f \int \frac{d^3p}{(2\pi)^3}\frac{p^i}{\omega_f}[(\omega_f -h_f)\delta f_f(x,p)+(\omega_f -\bar {h}_f)\delta \bar{f}_f(x,p)],
 \end{equation} 
 where we use the fact that in the rest frame of the fluid, heat flow four-vector and fluid four-vector are perpendicular to each other, {\em i.e.} $Q_{\mu}u^{\mu}=0$ and the enthalpy per particle for the $f^{th}$  flavour is given by
 \begin{equation}
 	h_f=\frac{(\epsilon_f+P_f)}{n_{eq,f}},	
 \end{equation}
 where
 \begin{equation}
 	\epsilon_f=g_f\int \frac{d^3p}{(2\pi)^3}\omega_f f_{eq,f},
 \end{equation}
 \begin{equation}
 	P_f=\frac{g_f}{3}\int \frac{d^3p}{(2\pi)^3}\frac{p^2}{\omega_f}f_{eq,f}.
 \end{equation}
 In the Navier-Stokes equation, the heat flow is related to the gradients of temperature and pressure by
 \begin{equation}
 	Q_{\mu}=\kappa\left[\nabla_{\mu}T - \frac{T}{(\epsilon+P)}\nabla_{\mu}P\right],	
 \end{equation}
 where $\kappa$ represents the thermal conductivity for the medium and $\nabla_{\mu}=\partial_\mu-u_{\mu}u_{\nu}\partial^\nu$. 

Now, the spatial component of the heat flow four-vector at a finite magnetic field can be expressed \cite{greif2013heat} as
 \begin{equation}
 	Q^i=-\kappa^{ij}\left[\partial_jT-\frac{T}{(\epsilon+P)}\partial_jP\right].
 \end{equation}
 Here, $\kappa^{ij}$ represents the thermal conductivity tensor \cite{greif2013heat},
 \begin{equation}
 	\kappa^{ij}=\kappa_0\delta^{ij}+ \kappa_1 \epsilon^{ijk}b_k+\kappa_2b^ib^j.
 \end{equation}
 In the above equation, $\kappa_0$, $\kappa_1$ and $\kappa_2$ are the various components of the thermal conductivity tensor. In this work, the gradients of temperature and pressure are taken perpendicular to the magnetic field, so the $\kappa_2$ term will vanish. Therefore, the above equation evolves to
 \begin{equation}\label{J}
 	Q^i=-(\kappa \delta^{ij}+\kappa_H \epsilon^{ij})\left[\partial_jT-\frac{T}{(n_{eq,f} h_f)}\partial_jP\right].
 \end{equation}
 Here $\kappa_0=\kappa$ and $\kappa_1=\kappa_H$ are known as thermal conductivity and Hall-type thermal conductivity, respectively. One can find the expressions for $\kappa$ and $\kappa_H$ using equations (\ref{K}) and (\ref{J}). 

In order to calculate the infinitesimal change in the distribution function ($\delta f_f$), we start with the RBTE given in eq. (\ref{E}) with the help of eq. (\ref{F}), 
 \begin{equation}\label{m2}
 	\frac{\tau_f}{p^0}p^{\mu}\frac{\partial f_{eq,f}}{\partial x^{\mu}} +\beta f_{eq,f}(\Gamma_x v_x+\Gamma_y v_y+\Gamma_z v_z)+\tau_f qEv_x\frac{\partial f_f}{\partial p^0}- qB\tau_f(v_x\frac{\partial f_f}{\partial p_y}-v_y\frac{\partial f_f}{\partial p_x})=0.	
 \end{equation}
 Using eq. (\ref{G}) and considering $L=\frac{\tau_f}{p^0}p^{\mu}\frac{\partial f_{eq,f}}{\partial x^{\mu}}$, eq. (\ref{m2}) is reduced to
 \begin{multline}\label{L}
 	L-\beta f_{eq,f}\tau_fqEv_x+\beta f_{eq,f}(\Gamma_x v_x+\Gamma_y v_y+\Gamma_z v_z)-\frac{qB\tau_f\beta f_{eq,f}}{\omega_f}(v_x\Gamma_y-v_y\Gamma_x) \\ +\frac{\tau^2_fq^2BEv_y\beta f_{eq,f}}{\omega_f}=0.    
 \end{multline}
 Here $L$ can be calculated as
 \begin{multline}
 	L=\tau_f\beta f_{eq,f}\frac{(\omega_f-h_f)}{T}v_x\left( \partial^xT-\frac{T}{nh_f}\partial^xP\right) +\tau_f\beta f_{eq,f} \frac{(\omega_f-h_f)}{T}v_y\left( \partial^yT-\frac{T}{nh_f}\partial^yP\right) \\
 	+\tau_f\beta f_{eq,f}\left[ p^0\frac{DT}{T}-\frac{p^{\mu}p^{\alpha}}{p^0}\nabla_{\mu}u_{\alpha}+TD\left(\frac{\mu_f}{T}\right)\right]. 
 \end{multline}
 Now, solving the eq. \eqref{L}\footnote{ For detailed calculations see appendix B of \cite{Rath2021}}, 
 we get
 \begin{multline}\label{x12}
 	\Gamma_x=qE\tau_{f}\frac{(1-\omega^2_c \tau^2_{f})}{(1+\omega^2_c \tau^2_{f})}-\frac{\tau_{f}}{(1+\omega^2_c \tau^2_{f})}\frac{(\omega_f-h_f)}{T}\left( \partial^xT-\frac{T}{nh_f}\partial^xP\right)\\
 	-\frac{\omega_c\tau^2_{f}}{(1+\omega^2_c \tau^2_{f})}\frac{(\omega_f-h_f)}{T}\left( \partial^yT-\frac{T}{nh_f}\partial^yP\right),
 \end{multline}
 \begin{multline}\label{x13}
 	\Gamma_y=-\frac{2qE\omega_c\tau^2_{f}}{(1+\omega^2_c \tau^2_{f})}-\frac{\tau_{f}}{(1+\omega^2_c \tau^2_{f})}\frac{(\omega_f-h_f)}{T}\left( \partial^yT-\frac{T}{nh_f}\partial^yP\right)\\
 	+\frac{\omega_c\tau^2_{f}}{(1+\omega^2_c \tau^2_{f})}\frac{(\omega_f-h_f)}{T}\left( \partial^xT-\frac{T}{nh_f}\partial^xP\right),
 \end{multline}
\begin{equation}\label{x14}
 \Gamma_z=0.
\end{equation}
 Putting the values of  $\Gamma_x$, $\Gamma_y$ and $\Gamma_z$ in equation (\ref{F}), we get
 \begin{multline}\label{M}
 	(f_f-n_f n^{-1}_{eq,f}f_{eq,f})=\frac{2qE\tau_fv_x \beta}{(1+\omega^2_c \tau^2_{f})}f_{eq,f}(1-f_{eq,f})-\frac{2qE\omega_c\tau^2_f v_y \beta}{(1+\omega^2_c \tau^2_{f})}f_{eq,f}(1-f_{eq,f})\\
 	-\beta f_{eq,f}(1-f_{eq,f})\frac{\tau_{f}}{(1+\omega^2_c \tau^2_{f})}\frac{(\omega_f-h_f)}{T}\left[ v_x\left( \partial^xT-\frac{T}{nh_f}\partial^xP\right)+v_y\left( \partial^yT-\frac{T}{nh_f}\partial^yP\right)\right]\\-\beta f_{eq,f}(1-f_{eq,f})\frac{\omega_c\tau^2_{f}}{(1+\omega^2_c \tau^2_{f})}\frac{(\omega_f-h_f)}{T}\left[ v_x\left( \partial^yT-\frac{T}{nh_f}\partial^yP\right)-v_y\left( \partial^xT-\frac{T}{nh_f}\partial^xP\right)\right].	
 \end{multline}
 With the help of the same approach, we can reduce the LHS of the above equation (\ref{M}) to obtain 
 \begin{multline}
 	\delta f_f-g_f n^{-1}_{eq,f} f_{eq,f}\int_{p} \delta f_f=\frac{2qE\tau_fv_x \beta}{(1+\omega^2_c \tau^2_{f})}f_{eq,f}(1-f_{eq,f})-\frac{2qE\omega_c\tau^2_f v_y \beta}{(1+\omega^2_c \tau^2_{f})}f_{eq,f}(1-f_{eq,f})\\
 	-\beta f_{eq,f}(1-f_{eq,f})\frac{\tau_{f}}{(1+\omega^2_c \tau^2_{f})}\frac{(\omega_f-h_f)}{T}\left[ v_x\left( \partial^xT-\frac{T}{nh_f}\partial^xP\right)+v_y\left( \partial^yT-\frac{T}{nh_f}\partial^yP\right)\right]\\-\beta f_{eq,f}(1-f_{eq,f})\frac{\omega_c\tau^2_{f}}{(1+\omega^2_c \tau^2_{f})}\frac{(\omega_f-h_f)}{T}\left[ v_x\left( \partial^yT-\frac{T}{nh_f}\partial^yP\right)-v_y\left( \partial^xT-\frac{T}{nh_f}\partial^xP\right)\right].	
 \end{multline}
 After neglecting higher order terms $\left(\mathcal O(\delta f_f)^2\right)$, we get the solution of this equation as
 \begin{equation}
 	\delta f_f=\delta f^{(0)}_f +g_f n^{-1}_{eq,f} f_{eq,f}\int_{p \prime } \delta f^{(0)}_f,
 \end{equation}
 where \\
 \begin{multline}
 	\delta f^{(0)}_f=\frac{2qE\tau_fv_x \beta}{(1+\omega^2_c \tau^2_{f})}f_{eq,f}(1-f_{eq,f})-\frac{2qE\omega_c\tau^2_f v_y \beta}{(1+\omega^2_c \tau^2_{f})}f_{eq,f}(1-f_{eq,f})\\
 	-\beta f_{eq,f}(1-f_{eq,f})\frac{\tau_{f}}{(1+\omega^2_c \tau^2_{f})}\frac{(\omega_f-h_f)}{T}\left[ v_x\left( \partial^xT-\frac{T}{nh_f}\partial^xP\right)+v_y\left( \partial^yT-\frac{T}{nh_f}\partial^yP\right)\right]\\-\beta f_{eq,f}(1-f_{eq,f})\frac{\omega_c\tau^2_{f}}{(1+\omega^2_c \tau^2_{f})}\frac{(\omega_f-h_f)}{T}\left[ v_x\left( \partial^yT-\frac{T}{nh_f}\partial^yP\right)-v_y\left( \partial^xT-\frac{T}{nh_f}\partial^xP\right)\right].
 \end{multline}
So finally, we get the infinitesimal change in the quark distribution function ($\delta f_f$) (see appendix \ref{appendix B}) as
 \begin{multline}\label{x10}
 	\delta f_f=\frac{2qE\tau_fv_x \beta}{(1+\omega^2_c \tau^2_{f})}f_{eq,f}(1-f_{eq,f})-\frac{2qE\omega_c\tau^2_f v_y \beta}{(1+\omega^2_c \tau^2_{f})}f_{eq,f}(1-f_{eq,f})\\
 	-\beta f_{eq,f}(1-f_{eq,f})\frac{\tau_{f}}{(1+\omega^2_c \tau^2_{f})}\frac{(\omega_f-h_f)}{T}\left[ v_x\left( \partial^xT-\frac{T}{nh_f}\partial^xP\right)+v_y\left( \partial^yT-\frac{T}{nh_f}\partial^yP\right)\right]\\-\beta f_{eq,f}(1-f_{eq,f})\frac{\omega_c\tau^2_{f}}{(1+\omega^2_c \tau^2_{f})}\frac{(\omega_f-h_f)}{T}\left[ v_x\left( \partial^yT-\frac{T}{nh_f}\partial^yP\right)-v_y\left( \partial^xT-\frac{T}{nh_f}\partial^xP\right)\right]\\+ g_f n^{-1}_{eq,f} f_{eq,f}\int_{ p\prime} \biggl[ \frac{2qE\tau_fv_x \beta}{(1+\omega^2_c \tau^2_{f})}f_{eq,f}(1-f_{eq,f})-\frac{2qE\omega_c\tau^2_f v_y \beta}{(1+\omega^2_c \tau^2_{f})}f_{eq,f}(1-f_{eq,f})\\
 	-\beta f_{eq,f}(1-f_{eq,f})\frac{\tau_{f}}{(1+\omega^2_c \tau^2_{f})}\frac{(\omega_f-h_f)}{T}\left[ v_x\left( \partial^xT-\frac{T}{nh_f}\partial^xP\right)+v_y\left( \partial^yT-\frac{T}{nh_f}\partial^yP\right)\right]\\-\beta f_{eq,f}(1-f_{eq,f})\frac{\omega_c\tau^2_{f}}{(1+\omega^2_c \tau^2_{f})}\frac{(\omega_f-h_f)}{T}\left[ v_x\left( \partial^yT-\frac{T}{nh_f}\partial^yP\right)-v_y\left( \partial^xT-\frac{T}{nh_f}\partial^xP\right)\right]\biggr] . 
 \end{multline}
 Similarly, for antiquarks, we get
 \begin{multline}\label{x11}
 	\delta \bar{f}_f=\frac{2qE\tau_{\bar{f}}v_x \beta}{(1+\omega^2_c \tau^2_{\bar{f}})}\bar{f}_{eq,f}(1-\bar{f}_{eq,f})-\frac{2qE\omega_c\tau^2_{\bar{f}} v_y \beta}{(1+\omega^2_c \tau^2_{\bar{f}})}\bar{f}_{eq,f}(1-\bar{f}_{eq,f})\\
 	-\beta \bar{f}_{eq,f}(1-\bar{f}_{eq,f})\frac{\tau_{\bar{f}}}{(1+\omega^2_c \tau^2_{\bar{f}})}\frac{(\omega_f-h_{\bar{f}})}{T}\left[ v_x\left( \partial^xT-\frac{T}{nh_f}\partial^xP\right)+v_y\left( \partial^yT-\frac{T}{nh_f}\partial^yP\right)\right]\\-\beta \bar{f}_{eq,f}(1-\bar{f}_{eq,f})\frac{\omega_c\tau^2_{\bar{f}}}{(1+\omega^2_c \tau^2_{\bar{f}})}\frac{(\omega_f-h_{\bar{f}})}{T}\left[ v_x\left( \partial^yT-\frac{T}{nh_f}\partial^yP\right)-v_y\left( \partial^xT-\frac{T}{nh_f}\partial^xP\right)\right]\\+ g_f n^{-1}_{eq,f} \bar{f}_{eq,f}\int_{p \prime }\biggl[  \frac{2qE\tau_{\bar{f}}v_x \beta}{(1+\omega^2_c \tau^2_{\bar{f}})}\bar{f}_{eq,f}(1-\bar{f}_{eq,f})-\frac{2qE\omega_c\tau^2_{\bar{f}} v_y \beta}{(1+\omega^2_c \tau^2_{\bar{f}})}\bar{f}_{eq,f}(1-\bar{f}_{eq,f})\\
 	-\beta \bar{f}_{eq,f}(1-\bar{f}_{eq,f})\frac{\tau_{\bar{f}}}{(1+\omega^2_c \tau^2_{\bar{f}})}\frac{(\omega_f-h_{\bar{f}})}{T}\left[ v_x\left( \partial^xT-\frac{T}{nh_f}\partial^xP\right)+v_y\left( \partial^yT-\frac{T}{nh_f}\partial^yP\right)\right]\\-\beta \bar{f}_{eq,f}(1-\bar{f}_{eq,f})\frac{\omega_c\tau^2_{\bar{f}}}{(1+\omega^2_c \tau^2_{\bar{f}})}\frac{(\omega_f-h_{\bar{f}})}{T}\left[ v_x\left( \partial^yT-\frac{T}{nh_f}\partial^yP\right)-v_y\left( \partial^xT-\frac{T}{nh_f}\partial^xP\right)\right]\biggr] .	
 \end{multline}
 Now, replacing the values of $\delta f_f$ and $\delta {\bar{f}}_f$ in equation (\ref{K}) and comparing it with equation (\ref{J}), we get the final expressions for thermal conductivity ($\kappa^{BGK}$) and Hall-type thermal conductivity ($\kappa^{BGK}_{H}$) under BGK model in weak magnetic field as
 \begin{multline}\label{x19}
 	\kappa^{BGK}=\frac{\beta^2}{6\pi^2}\sum_{f}g_f\int dp \frac{p^4}{\omega^2_f}\Biggl[ (\omega_f-h_f)^2\left(\frac{\tau_{f}}{1+\omega^2_c \tau^2_{f}}\right)f_{eq,f}(1-f_{eq,f})\\ +(\omega_f-h_{\bar{f}})^2\left( \frac{\tau_{\bar{f}}}{1+\omega^2_c\tau^2_{\bar{f}}}\right)\bar{f}_{eq,f}(1-\bar{f}_{eq,f})\Biggr]+\frac{\beta^2}{\sqrt{3}}\sum_{f}g^2_f  n^{-1}_{eq,f}\int_{ p} \frac{p}{\omega_f}(\omega_f-h_f)f_{eq,f}\\ \times \int_{ p\prime}\frac{p\prime}{\omega_f}\left[(\omega_f-h_f)\left(\frac{\tau_{f}}{1+\omega^2_c \tau^2_{f}}\right)f_{eq,f}(1-f_{eq,f})+(\omega_f-h_{\bar{f}})\left( \frac{\tau_{\bar{f}}}{1+\omega^2_c\tau^2_{\bar{f}}}\right)\bar{f}_{eq,f}(1-\bar{f}_{eq,f})\right],
 \end{multline}
 \begin{multline}\label{x20}
 	\kappa^{BGK}_{H}=\frac{\beta^2}{6\pi^2}\sum_{f}g_f\int dp \frac{p^4}{\omega^2_f}\Biggl[(\omega_f-h_f)^2\left(\frac{\omega_c\tau^2_{f}}{1+\omega^2_c \tau^2_{f}}\right)f_{eq,f}(1-f_{eq,f})\\+(\omega_f-h_{\bar{f}})^2\left( \frac{\omega_c\tau^2_{\bar{f}}}{1+\omega^2_c\tau^2_{\bar{f}}}\right)\bar{f}_{eq,f}(1-\bar{f}_{eq,f})\Biggr]+ \frac{\beta^2}{\sqrt{3}}\sum_{f}g^2_f  n^{-1}_{eq,f}\int_{ p} \frac{p}{\omega_f}(\omega_f-h_f)f_{eq,f}\\  \times  \int_{ p\prime}\frac{p\prime}{\omega_f} \left[(\omega_f-h_f)\left(\frac{\omega_c\tau^2_{f}}{1+\omega^2_c \tau^2_{f}}\right)f_{eq,f}(1-f_{eq,f})+(\omega_f-h_{\bar{f}})\left( \frac{\omega_c\tau^2_{\bar{f}}}{1+\omega^2_c\tau^2_{\bar{f}}}\right)\bar{f}_{eq,f}(1-\bar{f}_{eq,f})\right]. 	
 \end{multline}
 Here, we can write the expressions of $\kappa^{BGK}$ and $\kappa^{BGK}_{H}$ as an addition of two terms as
 \begin{equation}
 	\kappa^{BGK}=\kappa^{RTA}+\kappa^{Corr},
 \end{equation}
 \begin{equation}
 	\kappa^{BGK}_{H}=\kappa^{RTA}_{H}+\kappa^{Corr}_{H},
 \end{equation}
 where
 \begin{multline}
 	\kappa^{RTA}=\frac{\beta^2}{6\pi^2}\sum_{f}g_f\int dp \frac{p^4}{\omega^2_f}\Biggl[ (\omega_f-h_f)^2\left(\frac{\tau_{f}}{1+\omega^2_c \tau^2_{f}}\right)f_{eq,f}(1-f_{eq,f})\\
 	+(\omega_f-h_{\bar{f}})^2\left( \frac{\tau_{\bar{f}}}{1+\omega^2_c\tau^2_{\bar{f}}}\right)\bar{f}_{eq,f}(1-\bar{f}_{eq,f})\Biggl],
 \end{multline}
 \begin{multline}
 	\kappa^{corr}=\frac{\beta^2}{\sqrt{3}}\sum_{f}g^2_f  n^{-1}_{eq,f}\int_{ p} \frac{p}{\omega_f}(\omega_f-h_f)f_{eq,f} \times \int_{ p\prime}\frac{p\prime}{\omega_f}\Biggl[(\omega_f-h_f)\left(\frac{\tau_{f}}{1+\omega^2_c \tau^2_{f}}\right)f_{eq,f}(1-f_{eq,f})\\ +(\omega_f-h_{\bar{f}})\left( \frac{\tau_{\bar{f}}}{1+\omega^2_c\tau^2_{\bar{f}}}\right)\bar{f}_{eq,f}(1-\bar{f}_{eq,f})\Biggl],
 \end{multline}
 \begin{multline}
 	\kappa^{RTA}_{H}=\frac{\beta^2}{6\pi^2}\sum_{f}g_f\int dp \frac{p^4}{\omega^2_f}\Biggl[(\omega_f-h_f)^2\left(\frac{\omega_c\tau^2_{f}}{1+\omega^2_c \tau^2_{f}}\right)f_{eq,f}(1-f_{eq,f})\\
 	+(\omega_f-h_{\bar{f}})^2\left( \frac{\omega_c\tau^2_{\bar{f}}}{1+\omega^2_c\tau^2_{\bar{f}}}\right)\bar{f}_{eq,f}(1-\bar{f}_{eq,f})\Biggl],	
 \end{multline}
 \begin{multline}
 	\kappa^{Corr}_{H}=\frac{\beta^2}{\sqrt{3}}\sum_{f}g^2_f  n^{-1}_{eq,f}\int_{ p} \frac{p}{\omega_f}(\omega_f-h_f)f_{eq,f}  \times  \int_{ p\prime}\frac{p\prime}{\omega_f} \Biggl[(\omega_f-h_f)\left(\frac{\omega_c\tau^2_{f}}{1+\omega^2_c \tau^2_{f}}\right)f_{eq,f}(1-f_{eq,f})\\+(\omega_f-h_{\bar{f}})\left( \frac{\omega_c\tau^2_{\bar{f}}}{1+\omega^2_c\tau^2_{\bar{f}}}\right)\bar{f}_{eq,f}(1-\bar{f}_{eq,f})\Biggl]. 	
 \end{multline}
 
 \section{Results and discussions}\label{xxc}
\begin{figure}[h]
 	\begin{subfigure}[b]{0.5\linewidth}
 	\includegraphics[width=1\textwidth]{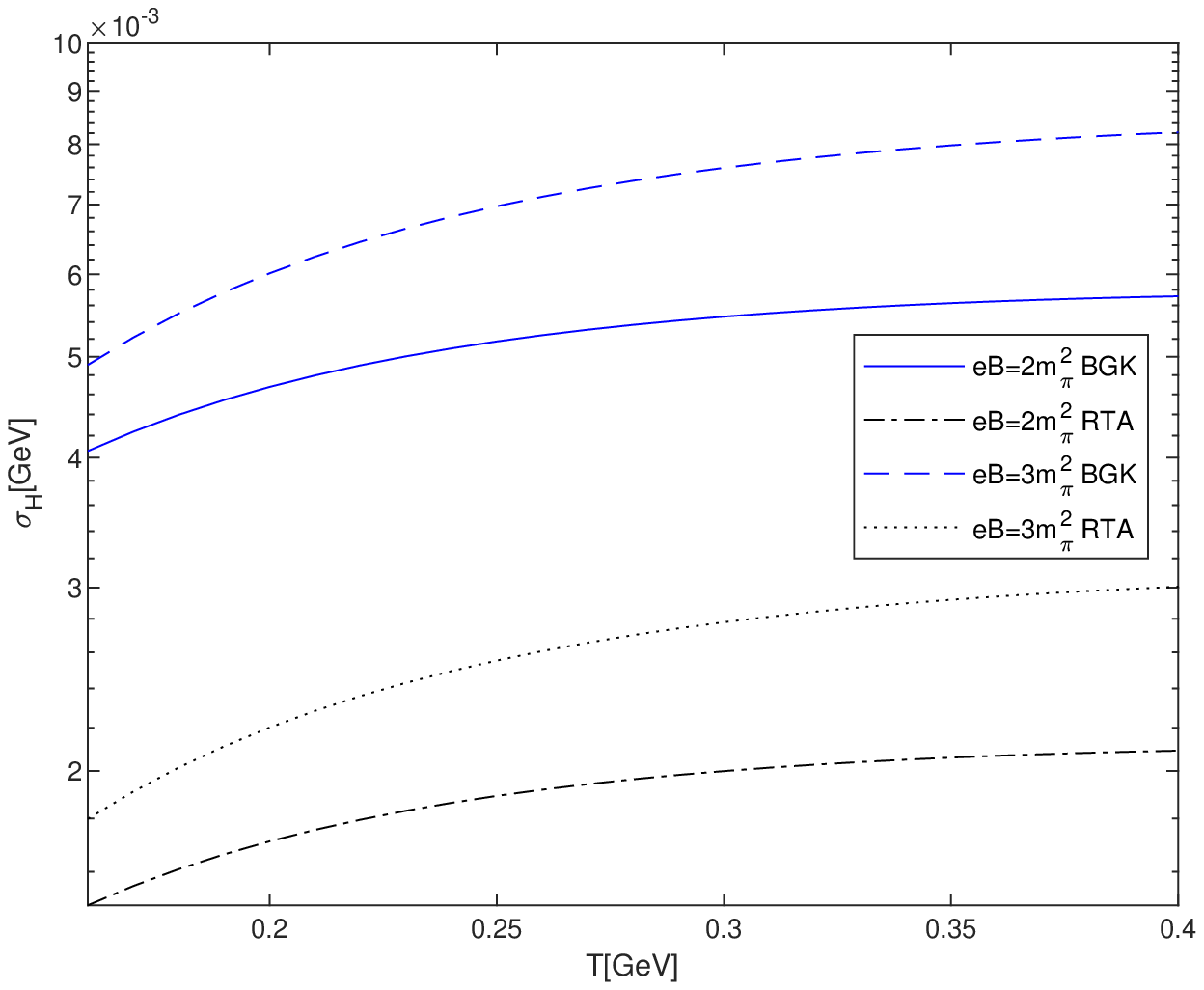}
 	\caption{}\label{a03}
 \end{subfigure}
 	\hfill
 	\begin{subfigure}[b]{0.5\linewidth}
 	\includegraphics[width=1\textwidth]{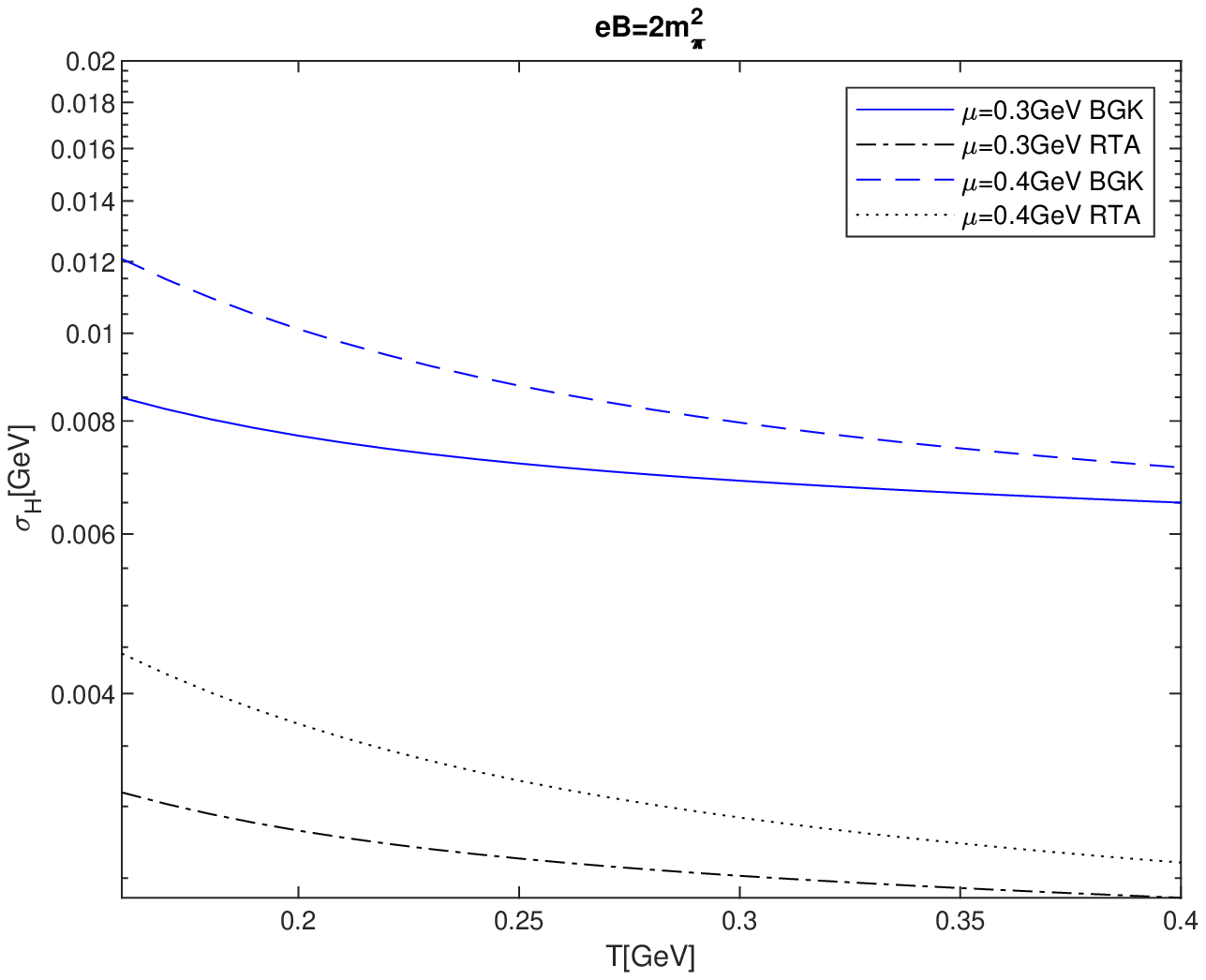}
 	\caption{}\label{a04}
 \end{subfigure}
 	\caption{The variation of the Hall conductivity ($\sigma_{H}$) with temperature, (a) comparison with RTA model result in the presence of weak magnetic field and (b) comparison with RTA model result in the presence of finite chemical potential.}\label{a002}
 \end{figure}

The charge transport coefficients ($\sigma_{el}$, $\sigma_{H}$) and heat transport coefficients ($\kappa$, $\kappa_{H}$) are estimated in the presence of weak magnetic field within the framework 
of the BGK model by considering the current masses of the quarks. 
 \begin{figure}[h]
 \begin{subfigure}[b]{0.5\linewidth}
 		\includegraphics[width=1\textwidth]{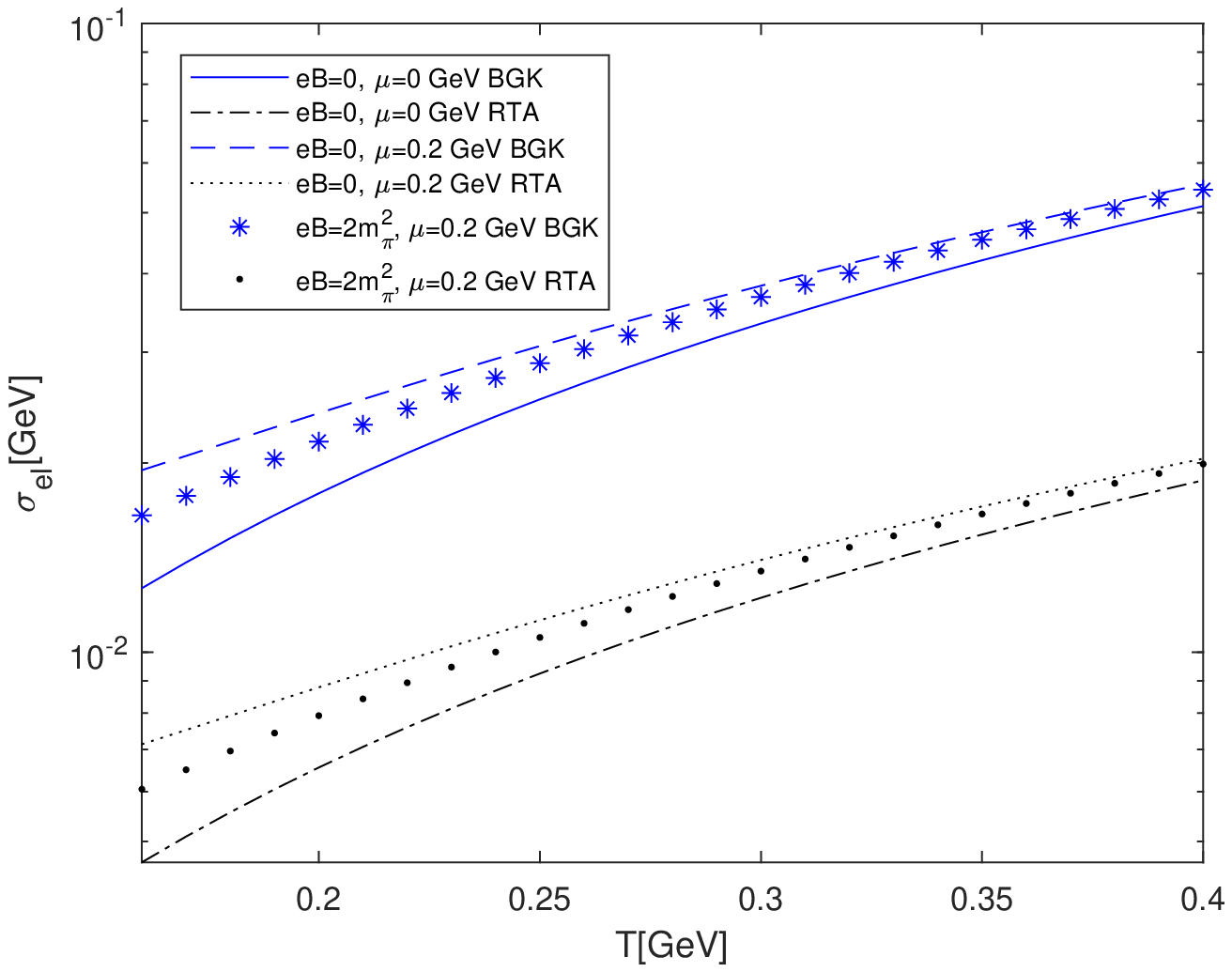}
 		\caption{}\label{a01}
 	\end{subfigure}
 	\hfill
 	\begin{subfigure}[b]{0.5\linewidth}
 		\includegraphics[width=1\textwidth]{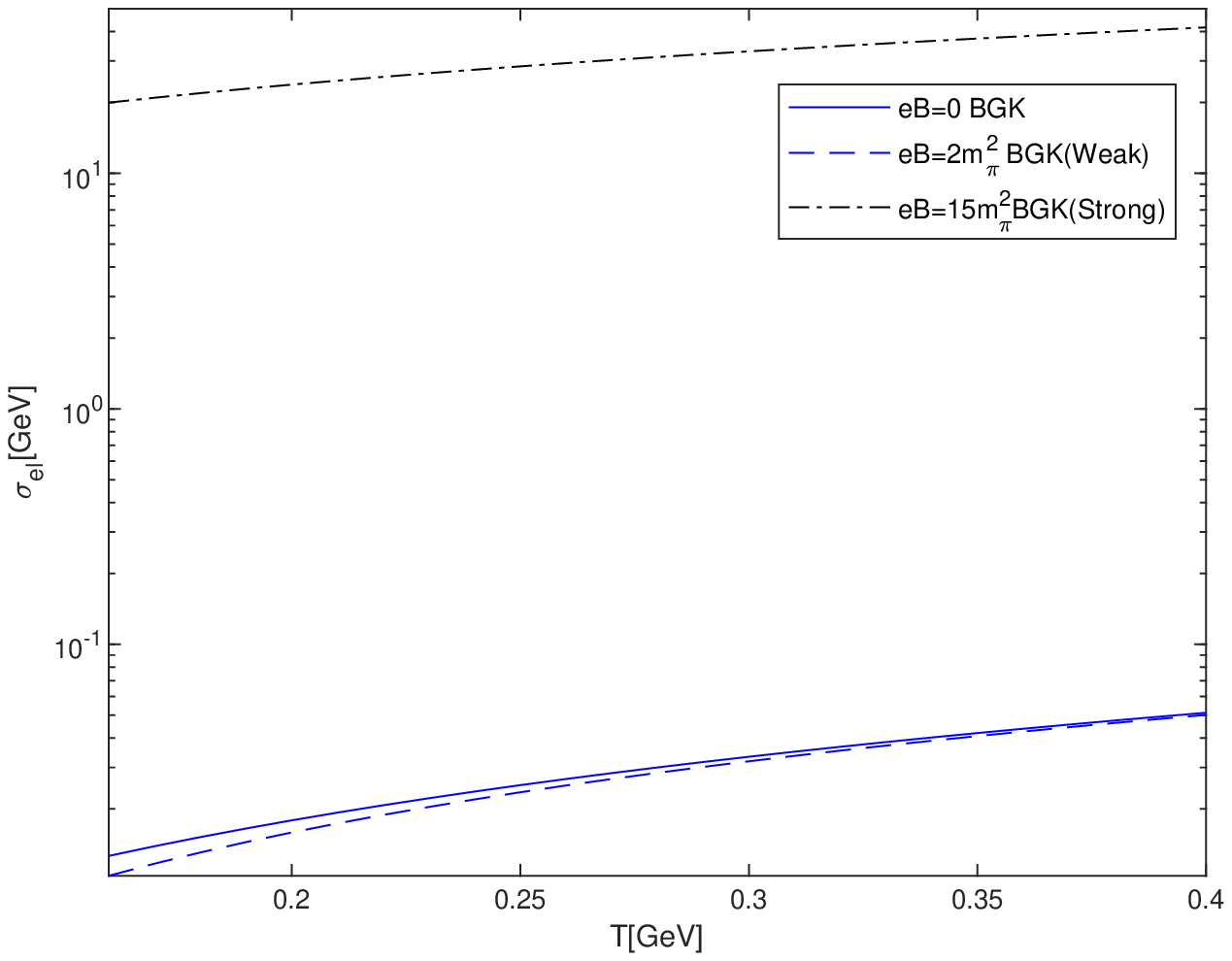}
 		\caption{}\label{a02}
 	\end{subfigure}
 	\caption{The variation of the electrical conductivity ($\sigma_{el}$) with temperature, (a) comparison with RTA model result (weak magnetic field) and (b) comparison with BGK model result (strong magnetic field).}\label{a001}
 \end{figure}
Figures \ref{a001} and \ref{a002} show the variations of $\sigma_{el}$ and $\sigma_{H}$ as function of  temperature, respectively. In figure \ref{a01}, we compare the estimated results with the RTA model in the weak magnetic field regime. We have found that due to this new BGK collision term, there is an increment in the charge transport phenomenon for the QCD medium. This is indicated by the large value of electrical conductivity. The ratio of ${\sigma^{BGK}_{el}}/{\sigma^{RTA}_{el}}$ comes out to be 2.72. In figure \ref{a02}, similar kind of comparison has been done with BGK model results for the strong magnetic field and it is observed that the weak magnetic field slightly reduces the electrical conductivity, contrary to the  enhancement of the same by the strong magnetic field. \\

On the other hand, in figure \ref{a002}, we have done a suchlike comparison of our estimated results of Hall conductivity ($\sigma_{H}$). As presented in figure \ref{a03}, it is found that the Hall conductivity shows increasing behaviour with temperature (T) for both RTA and BGK model at zero chemical potential ($\mu=0$), whereas figure \ref{a04} shows decreasing behaviour with temperature (T) at finite chemical potential. 
 \\

\begin{figure}[h]
 	\begin{subfigure}[b]{0.5\linewidth}
 	\includegraphics[width=1\textwidth]{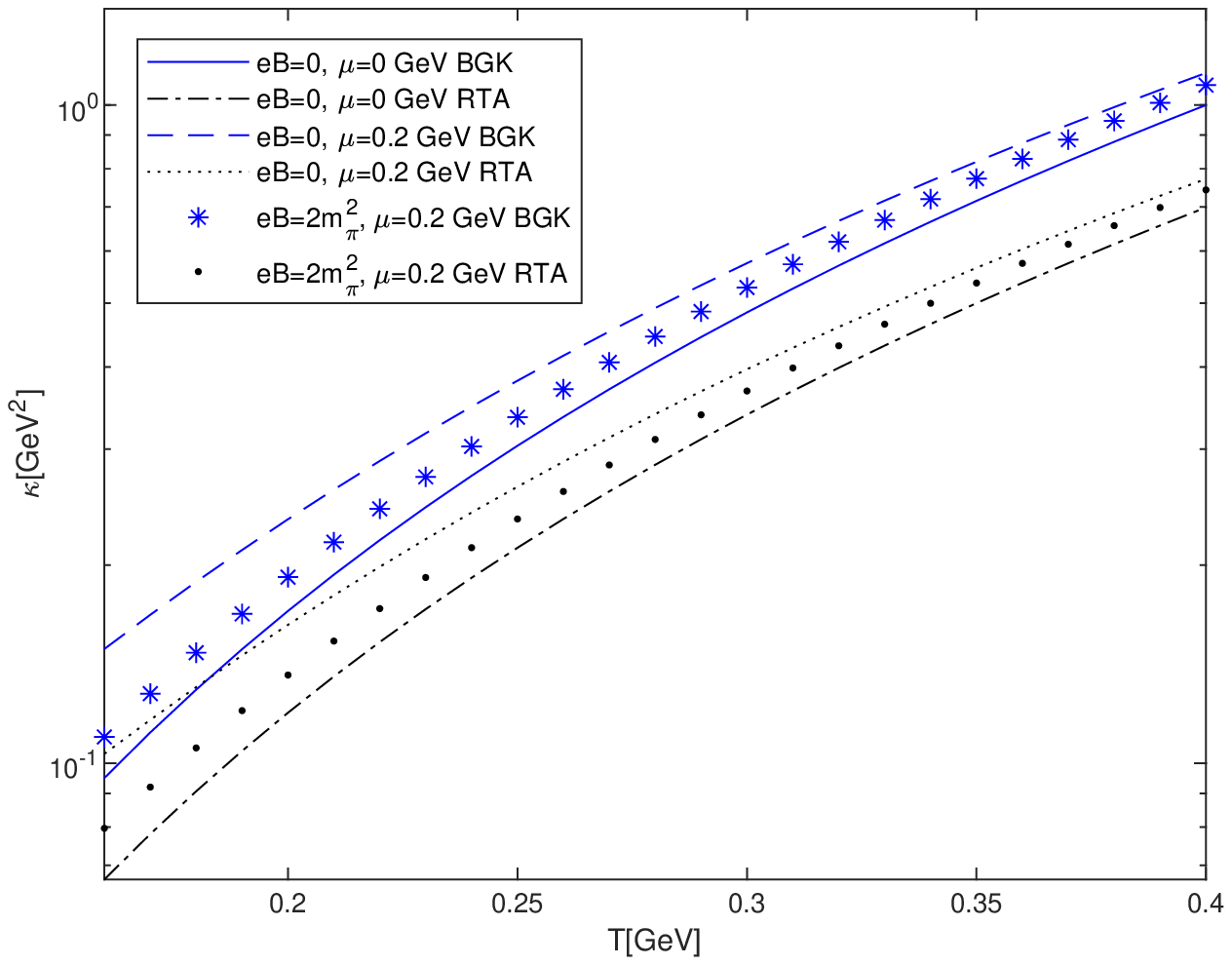}
 	\caption{}\label{a05}
 \end{subfigure}
 	\hfill
 	\begin{subfigure}[b]{0.5\linewidth}
 	\includegraphics[width=1\textwidth]{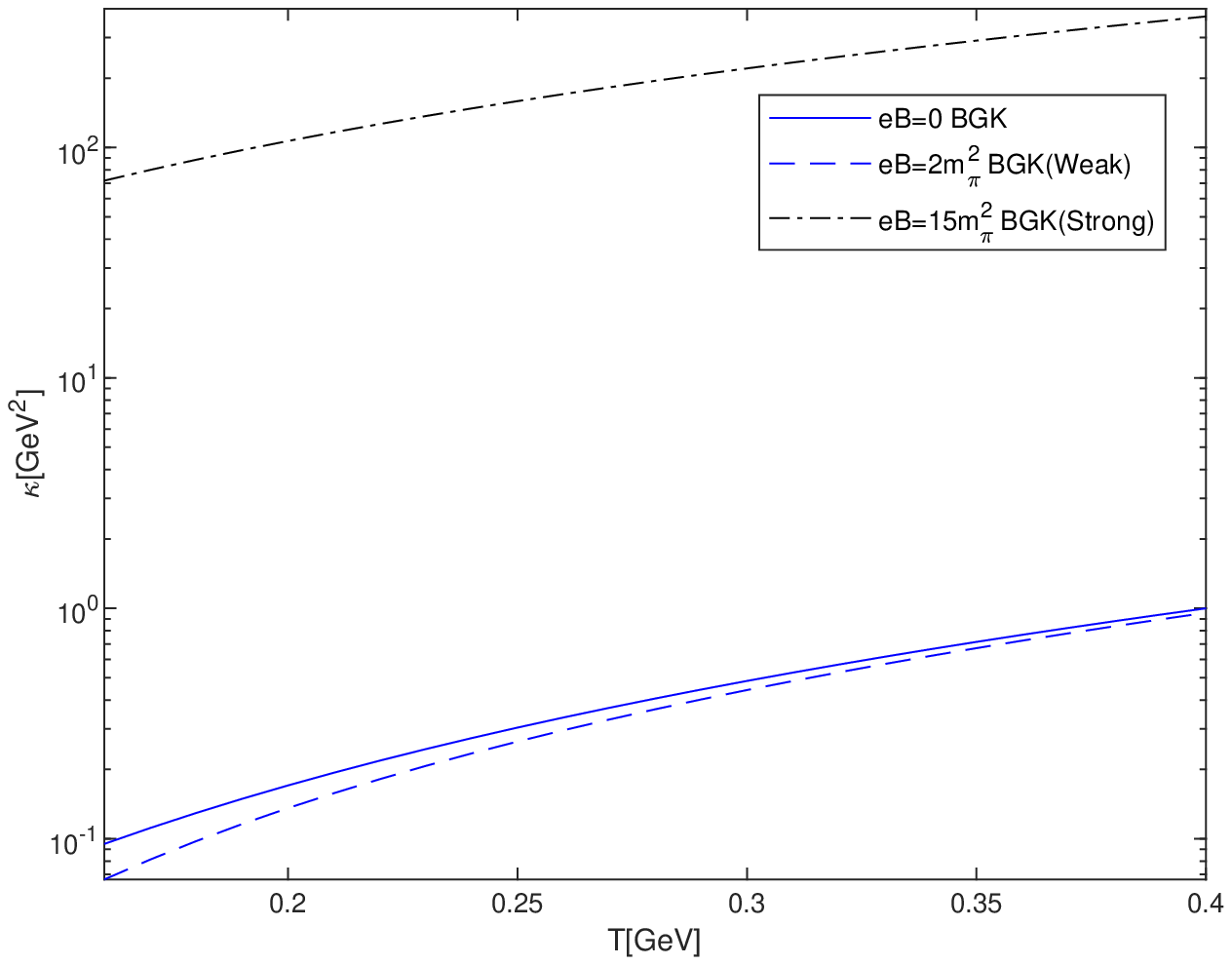}
 	\caption{}\label{a06}
 \end{subfigure}
 	\caption{The variation of the thermal conductivity ($\kappa$) with temperature, (a) comparison with RTA model result (weak magnetic field) and (b) comparison with BGK model result (strong magnetic field).}\label{a003}
 \end{figure}

\begin{figure}[H]
 	\begin{subfigure}[b]{0.5\linewidth}
 	\includegraphics[width=1\textwidth]{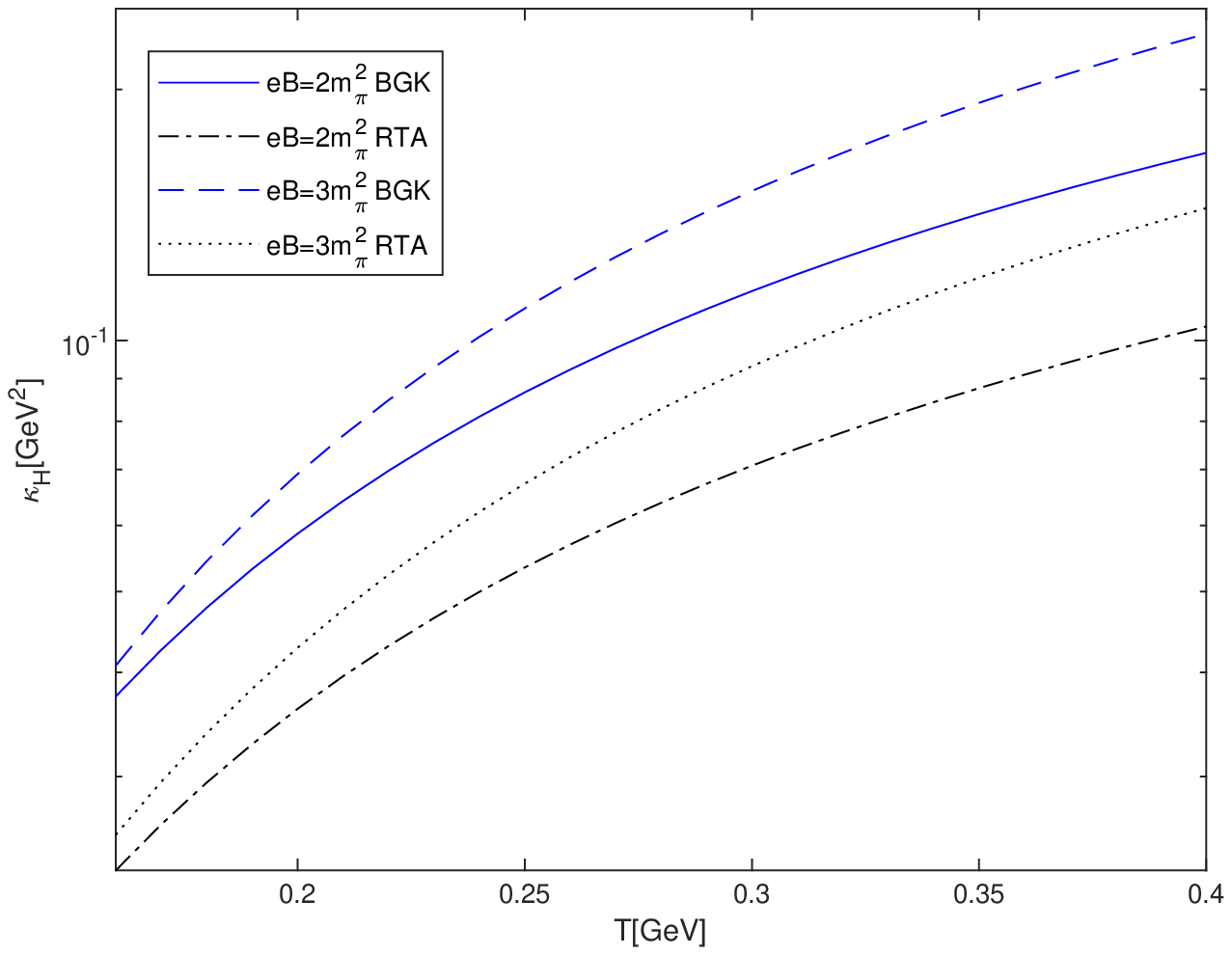}
 		\caption{}\label{a07}
 \end{subfigure}
 	\hfill
 	\begin{subfigure}[b]{0.5\linewidth}
 	\includegraphics[width=1\textwidth]{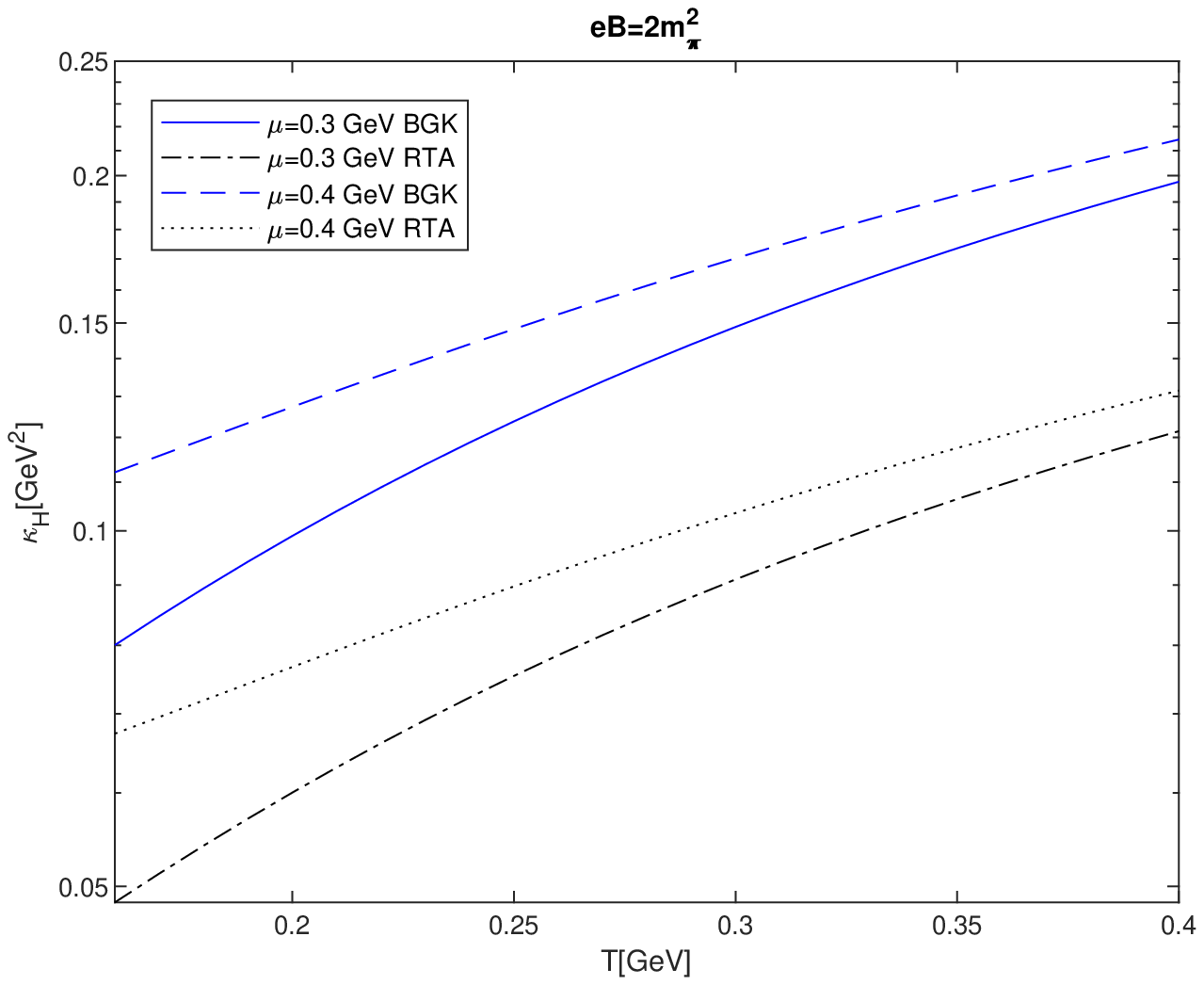}
 		\caption{}\label{a08}
 \end{subfigure}
 	\caption{The variation of the Hall-type thermal conductivity ($\kappa_{H}$) with temperature, (a) comparison with RTA model result in the presence of weak magnetic field and (b) comparison with RTA model result in the presence of finite chemical potential.}\label{a004}
 \end{figure}

Figures \ref{a003} and \ref{a004} show the variations of $\kappa$ and $\kappa_{H}$ as functions of temperature. To be more specific, figures \ref{a05} and \ref{a06} represent the comparison of our estimated results of thermal conductivity ($\kappa$) with RTA model results of weak magnetic field and BGK model results of strong magnetic field, respectively, whereas figures \ref{a07} and \ref{a08} show the comparison of our calculated results of $\kappa_{H}$ with RTA model results for weak magnetic field and for finite chemical potential, respectively. The ratio of ${\kappa^{BGK}}/{\kappa^{RTA}}$ comes out to be 1.42. Thus the above discussions imply that our collision integral is more sensitive to charge transport than heat transport. 


 \section{Quasiparticle model (QPM) and the transport coefficients}
 \subsection{Quasiparticle description of QGP}
Noninteracting quasiparticles in a thermal medium obtain some masses due to the interaction with the medium known as thermal mass. In the present context, we can consider QGP as a system of massive noninteracting quasiparticles in the quasiparticle model framework. This quasiparticle quark-gluon plasma (qQGP) model is widely used to describe the nonideal behaviour of QGP. In the case of a pure thermal medium, the quasiparticle masses of particles depend only on the temperature of the medium, but for a dense thermal medium, they also depend on chemical potential. The thermal mass (squared) of quark for a dense QCD medium is given \cite{braaten1992simple, peshier2002qcd} by 
 \begin{equation}\label{N}
 	\centering
 	m^2_{fT}=\frac{g^2T^2}{6}\left( 1+\frac{\mu^2_f}{\pi^2T^2}\right),
 \end{equation}
with $g=(4\pi\alpha_s)^{{1}/{2}}$. Here the chemical potentials for all the three quarks are set to be the same ($\mu=\mu_f$). 

Due to these temperature-dependent masses, there is a significant change in the transport phenomena of the given system and also in the corresponding transport coefficients. In the following section, our aim is to study the variations of electrical, Hall, thermal and Hall-type thermal conductivities with temperature considering the quasiparticle masses of the quarks. 

\subsection{Results and discussions}
\begin{figure}[ht]
	\begin{subfigure}{0.5\linewidth}
		\includegraphics[width=1\textwidth]{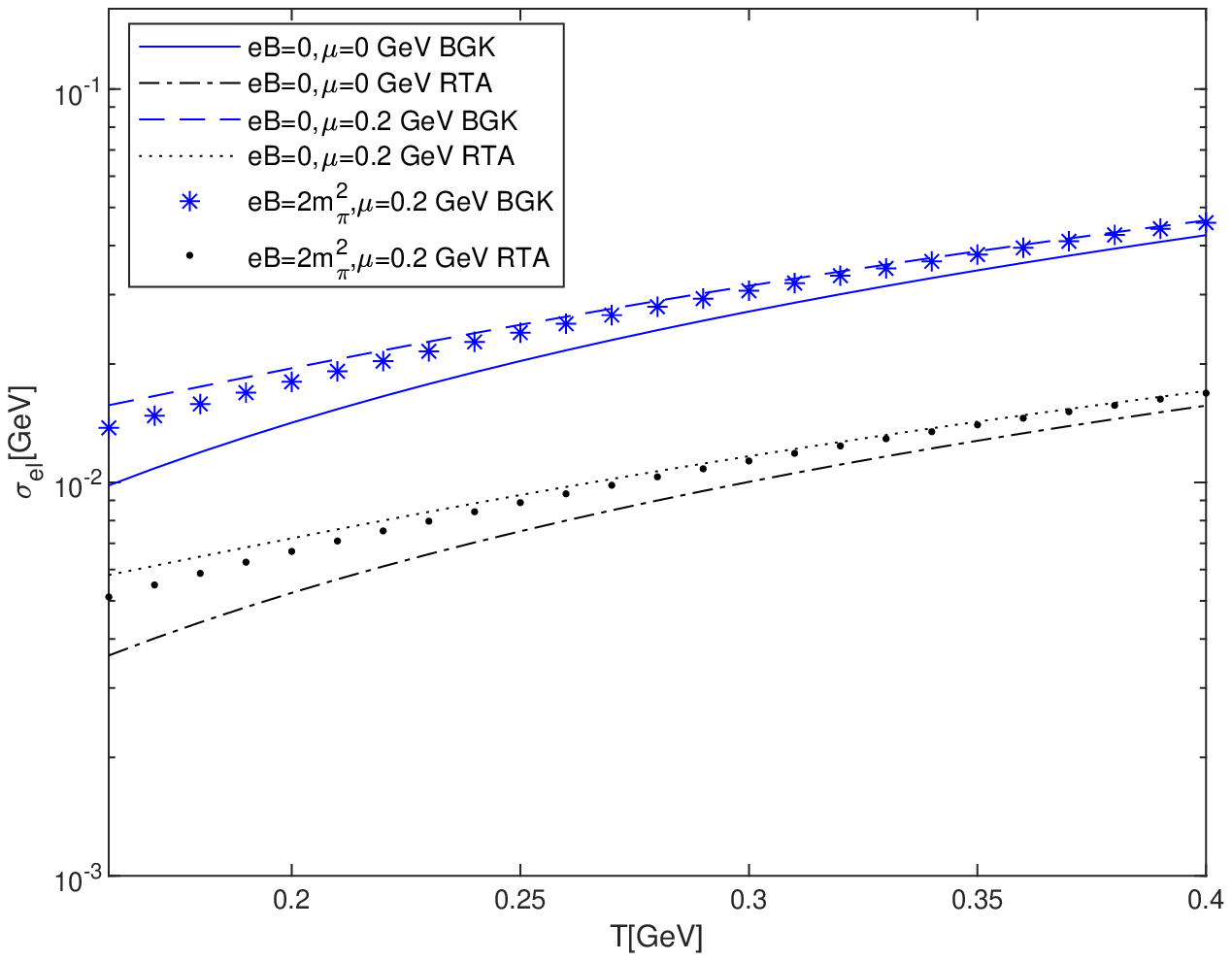}
		\caption{}\label{a09}
	\end{subfigure}
	\hfill
	\begin{subfigure}{0.5\linewidth}
		\includegraphics[width=1\textwidth]{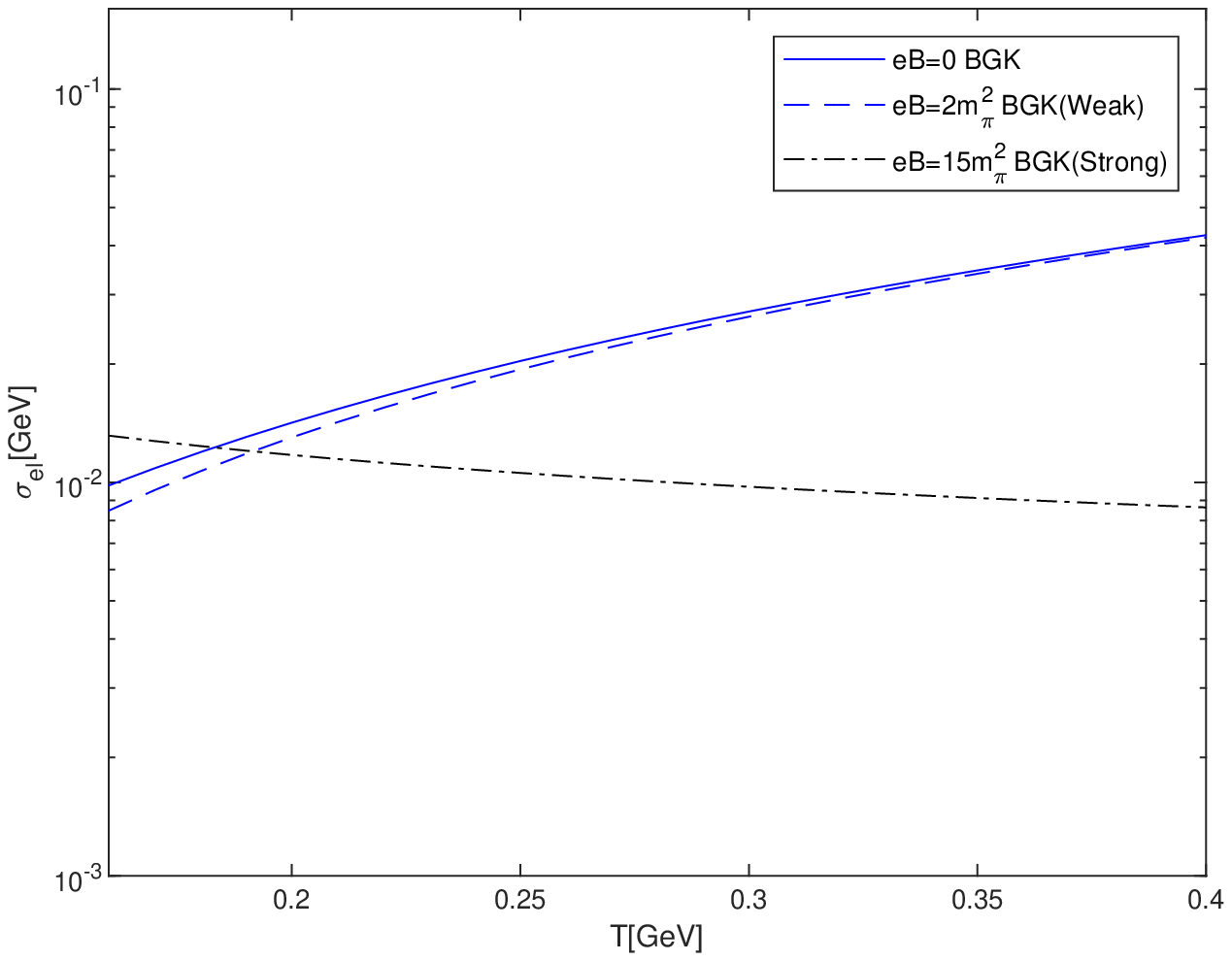}
		\caption{}\label{a010}
	\end{subfigure}
	\caption{The variation of the electrical conductivity ($\sigma_{el}$) with temperature, (a) comparison with RTA model result (weak magnetic field) and (b) comparison with BGK model result (strong magnetic field).}\label{a005}
\end{figure}

\begin{figure}[H]
	\begin{subfigure}{0.5\linewidth}
		\includegraphics[width=1\textwidth]{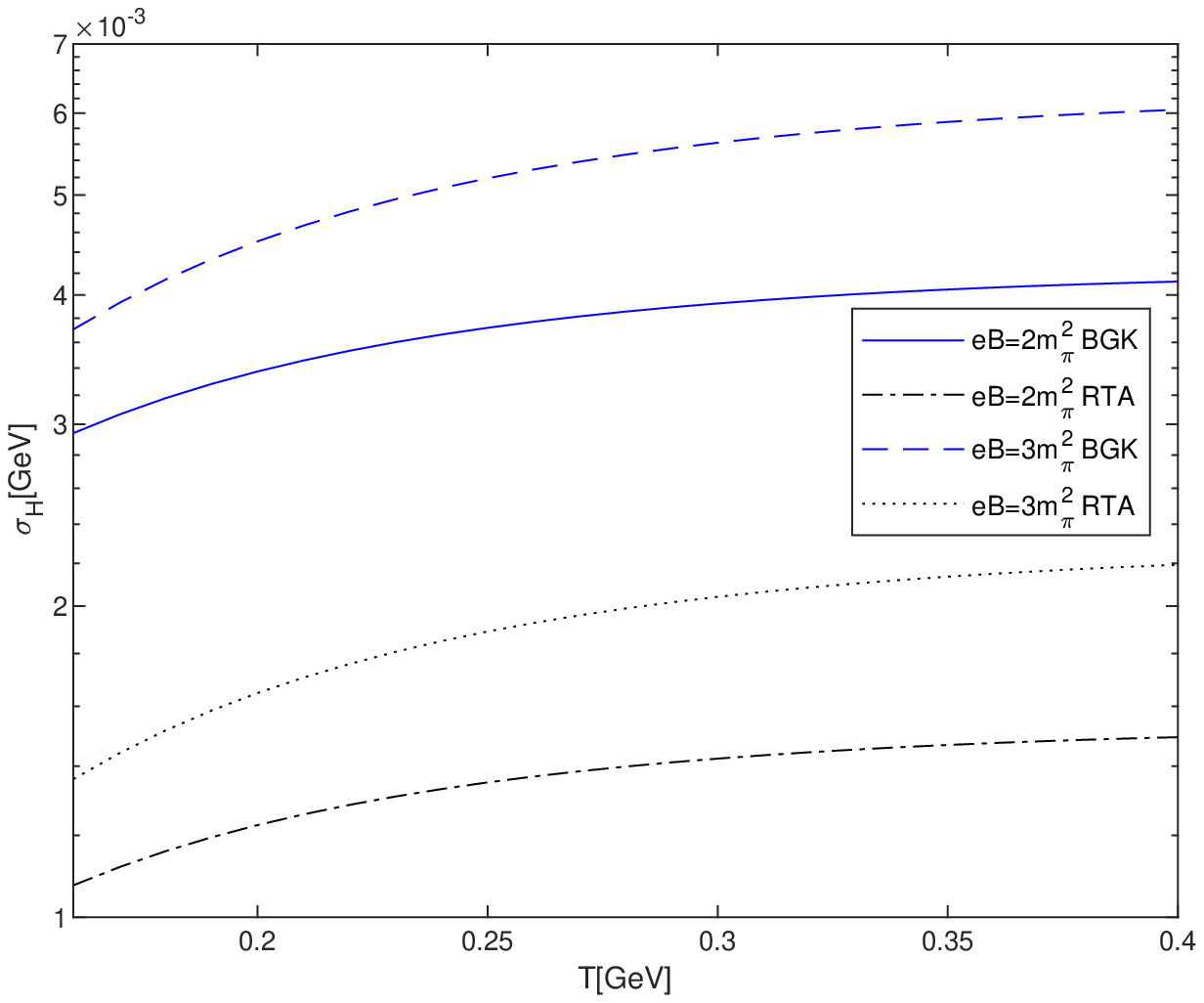}
		\caption{}\label{a011}
	\end{subfigure}
	\hfill
	\begin{subfigure}{0.5\linewidth}
		\includegraphics[width=1\textwidth]{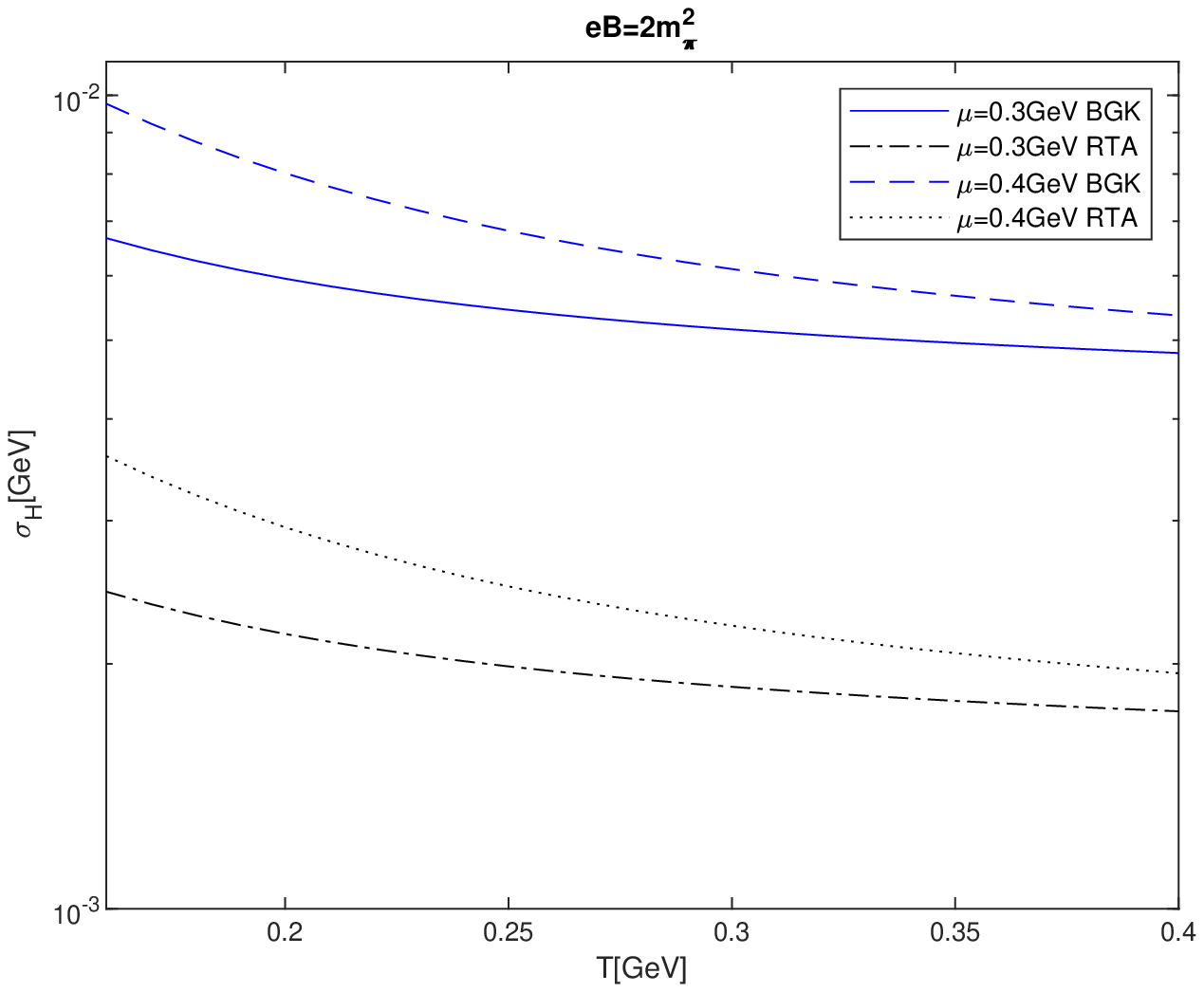}
		\caption{}\label{a012}
	\end{subfigure}
	\caption{The variation of the Hall conductivity ($\sigma_{H}$) with temperature, (a) comparison with RTA model result in the presence of weak magnetic field and (b) comparison with RTA model result in the presence of finite chemical potential.}\label{a006}
\end{figure}

\begin{figure}[h]
	\begin{subfigure}{0.5\linewidth}
		\includegraphics[width=1\textwidth]{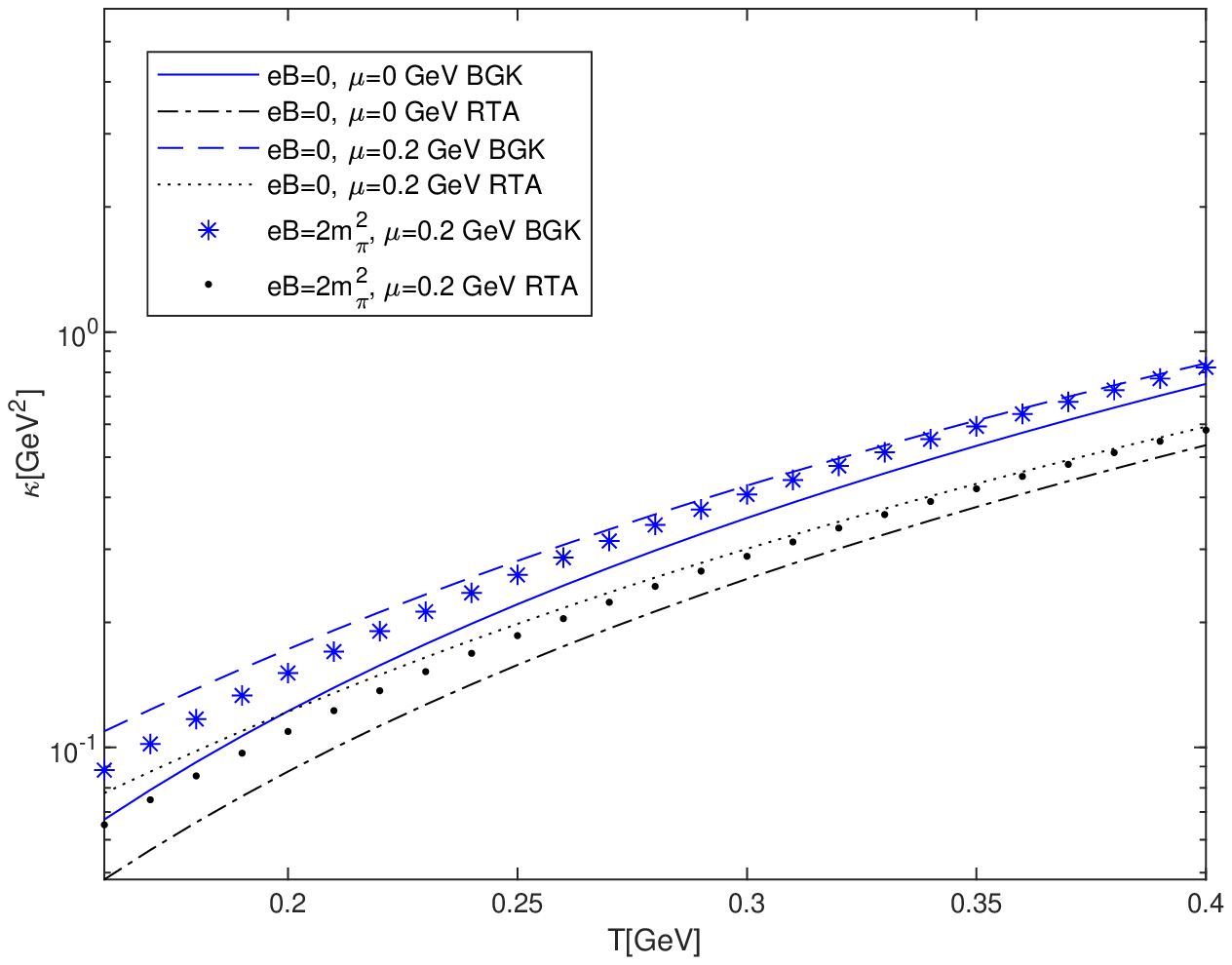}
		\caption{}\label{a013}
	\end{subfigure}
	\hfill
	\begin{subfigure}{0.5\linewidth}
		\includegraphics[width=1\textwidth]{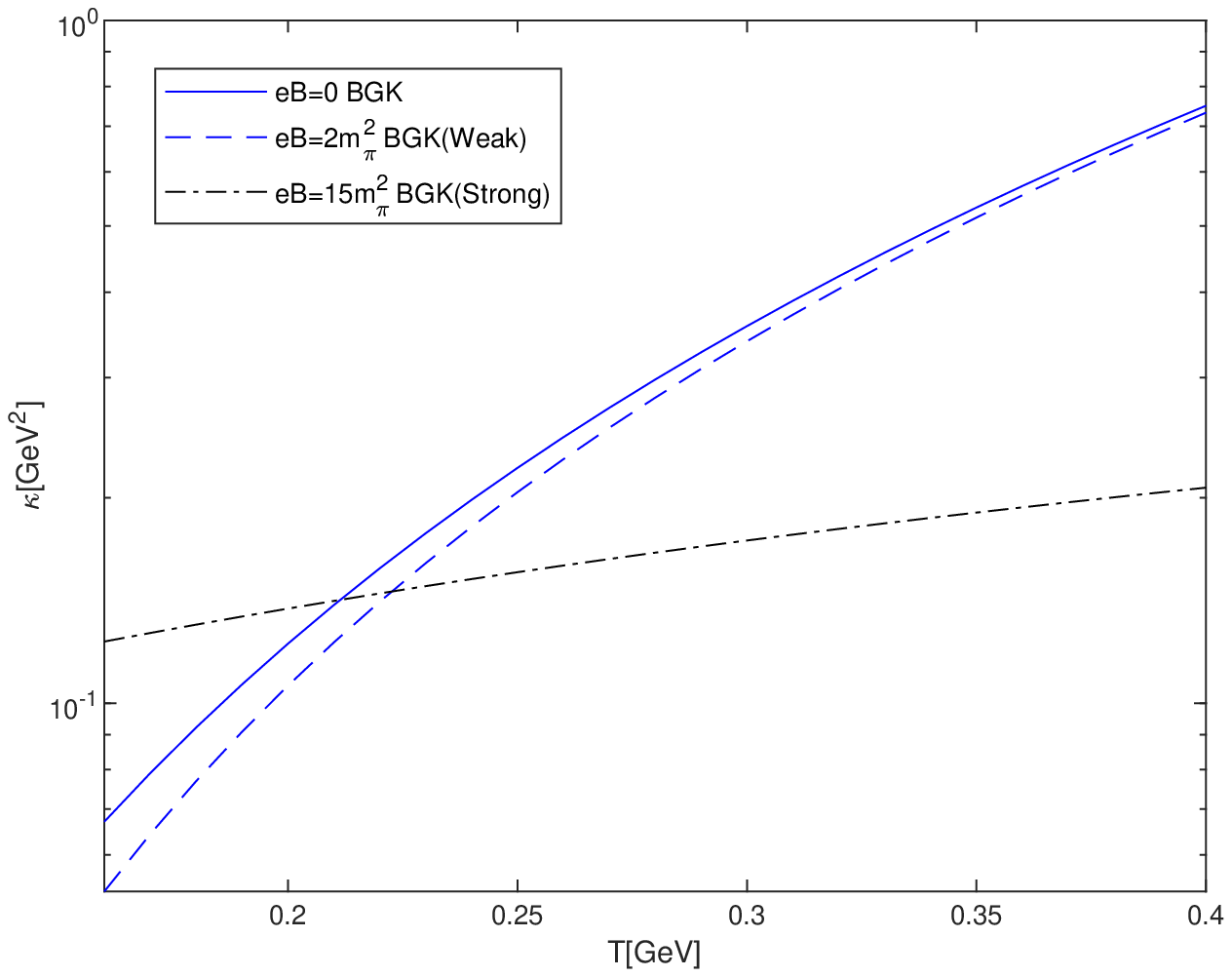}
		\caption{}\label{a014}
	\end{subfigure}
	\caption{The variation of the thermal conductivity ($\kappa$) with temperature, (a) comparison with RTA model result (weak magnetic field) and (b) comparison with BGK model result (strong magnetic field).}\label{a007}
\end{figure}

\begin{figure}[H]
	\begin{subfigure}{0.5\linewidth}
		\includegraphics[width=1\textwidth]{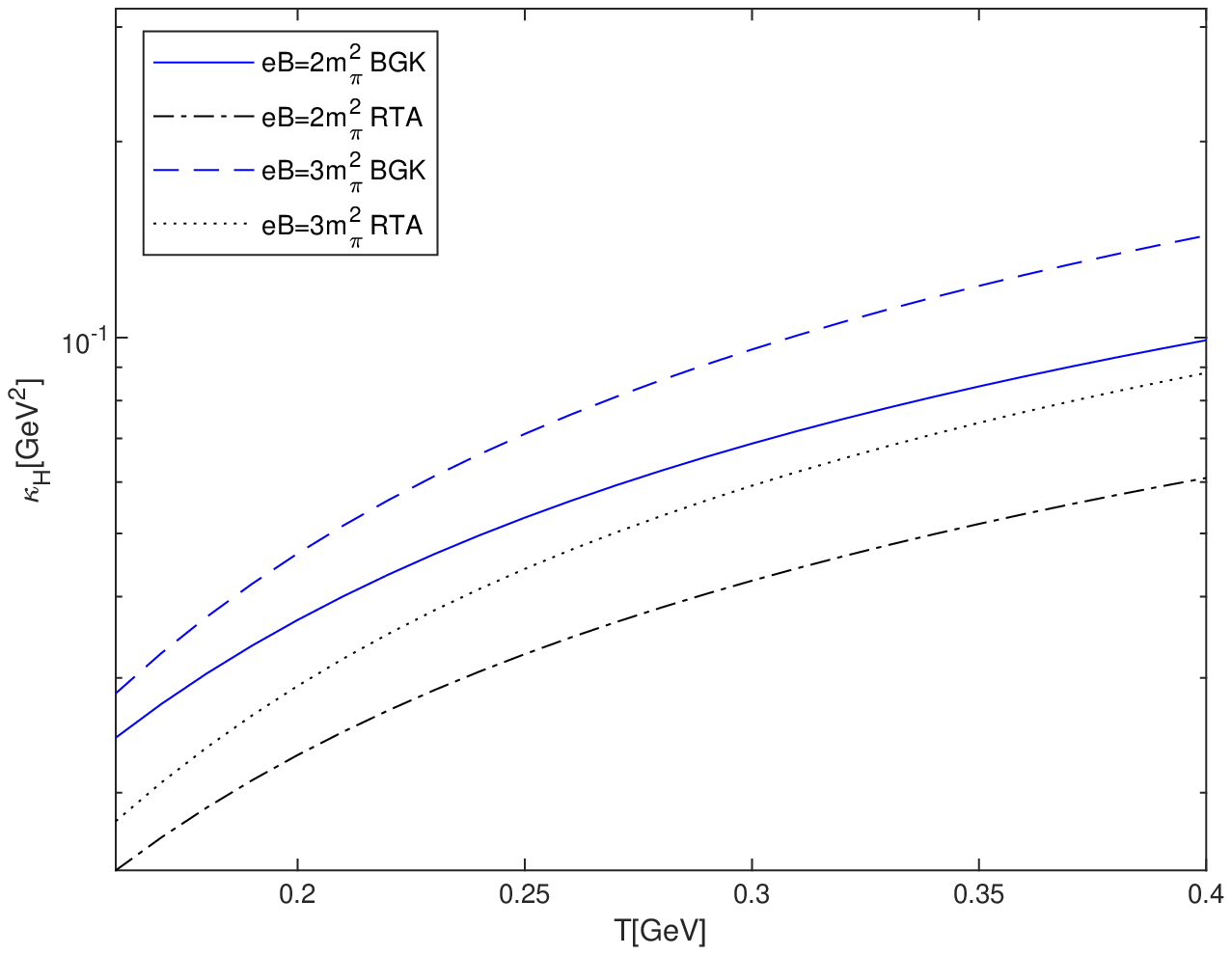}
		\caption{}\label{a015}
	\end{subfigure}
	\hfill
	\begin{subfigure}{0.5\linewidth}
		\includegraphics[width=1\textwidth]{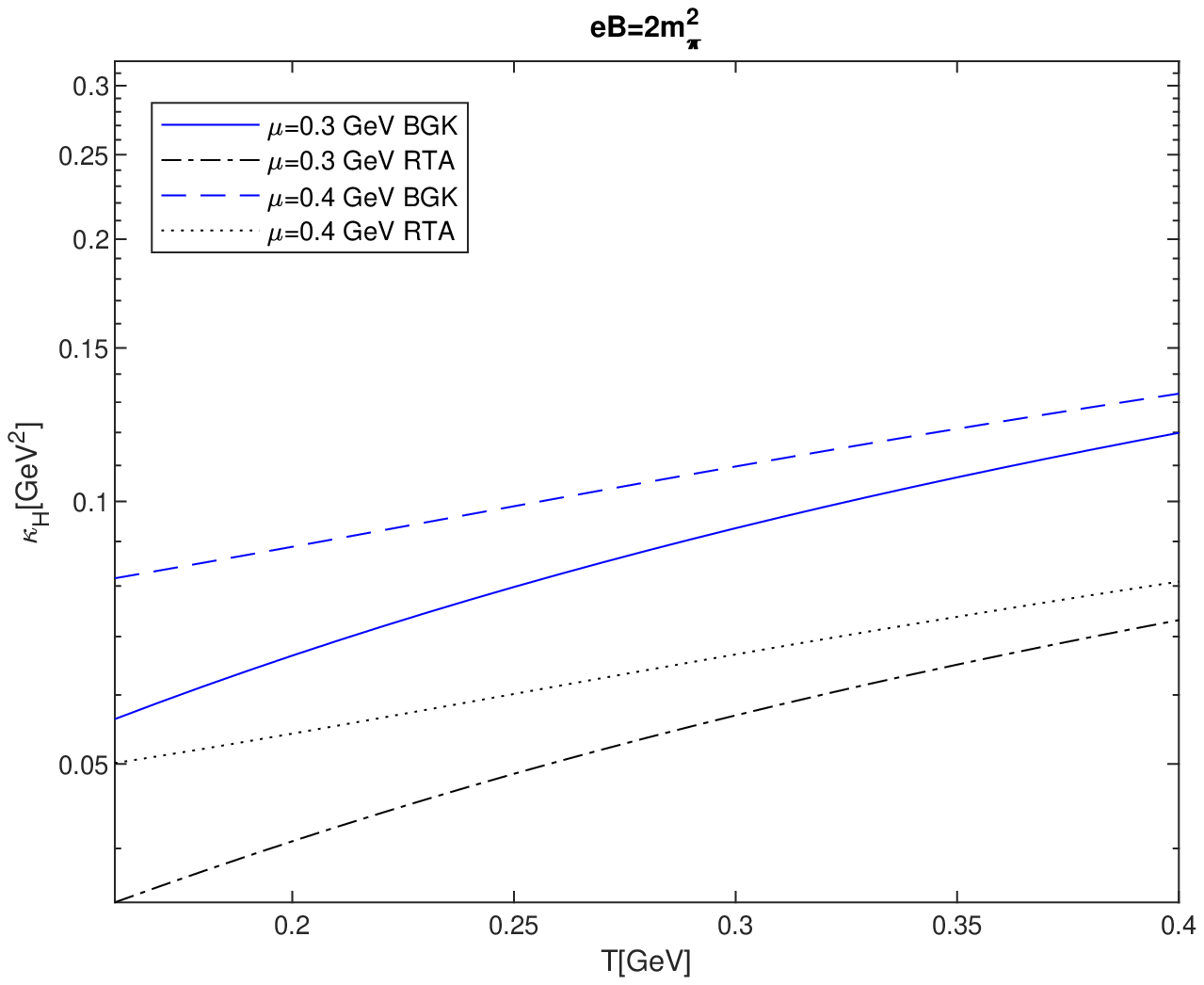}
		\caption{}\label{a016}
	\end{subfigure}
	\caption{The variation of the Hall-type thermal conductivity ($\kappa_{H}$) with temperature, (a) comparison with RTA model result in the presence of weak magnetic field and (b) comparison with RTA model result in the presence of finite chemical potential.}\label{a008}
\end{figure}

In this section, the charge and thermal coefficients were  reinvestigated  with the thermal mass as given by the equation $(\ref{N})$. Figures \ref{a005} and \ref{a006} show the variations of $\sigma_{el}$ and $\sigma_{H}$ with temperature, respectively. To be more specific, in figures \ref{a09} and \ref{a010}, we compare our results of electrical conductivity with the results of the RTA model (weak magnetic field) and BGK model (strong magnetic field). Similarly, figures \ref{a011} and \ref{a012} represent the comparison of Hall conductivity with the results of the RTA model in weak magnetic field and finite chemical potential, respectively. In figures \ref{a007} and \ref{a008}, same comparisons have been done for $\kappa$ and $\kappa_{H}$, respectively. In this model, all the transport coefficients depend on the distribution functions of quarks and gluons in the medium. The distribution functions are observed to be slightly modified due to the quasiparticle masses of quarks and gluons, which further causes change in the transport phenomena of the medium. In the calculation of conductivities and corresponding coefficients, we follow the similar methodology as done for the current quark masses but with  masses  being replaced by the medium-generated masses, {\em i.e.} quasiparticle masses. 

It is observed that the heat and charge transport phenomena slightly slow down due to the quasiparticle masses (comparatively heavier) than current quark masses because of the reduced mobility of carriers. As a result, we see a slightly higher values of $\sigma_{el}$, $\sigma_H$, $\kappa$ and $\kappa_H$ in the case of current quark masses. The ratios of ${\sigma^{BGK}_{el}}/{\sigma^{RTA}_{el}}$ and ${\kappa^{BGK}}/{\kappa^{RTA}}$ are observed to be 2.70 (figure \ref{a09}) and 1.39 (figure \ref{a013}), respectively. This shows that the collision integral is more sensitive to charge transport, irrespective of quark masses. 

\section{Applications}
\subsection{Wiedemann-Franz law and Lorenz number}
This law states that the ratio of thermal conductivity ($\kappa$) to electrical 
conductivity ($\sigma_{el}$) is directly proportional to the temperature (T) through 
a proportionality factor known as the Lorenz number ($L$). 
\begin{equation}
	\frac{\kappa}{\sigma_{el}}=LT
.\end{equation}
\begin{figure}[H]
	\begin{subfigure}{0.5\linewidth}
		\includegraphics[width=1\textwidth]{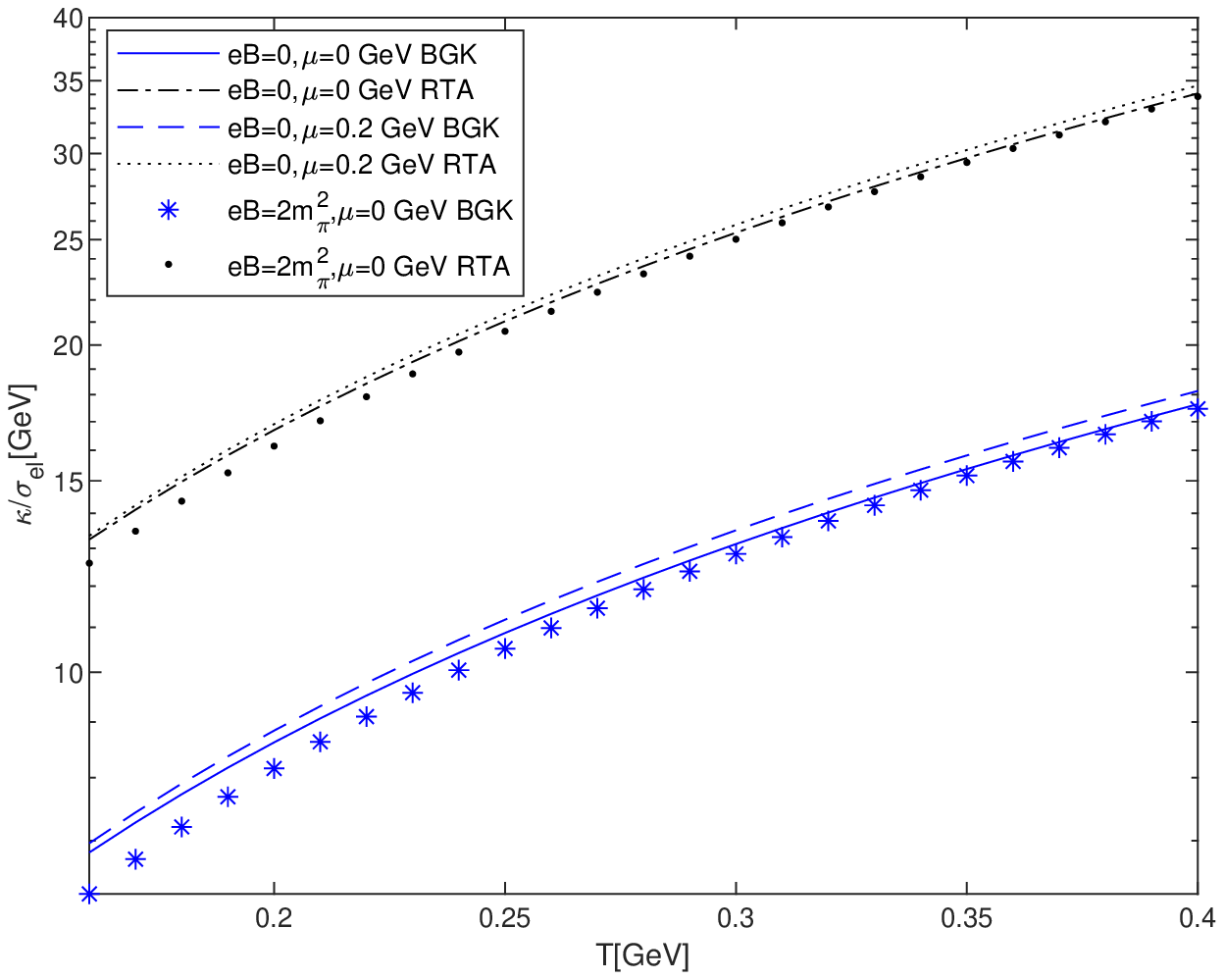}
		\caption{}\label{a017}
	\end{subfigure}
	\hfill
	\begin{subfigure}{0.5\linewidth}
		\includegraphics[width=1\textwidth]{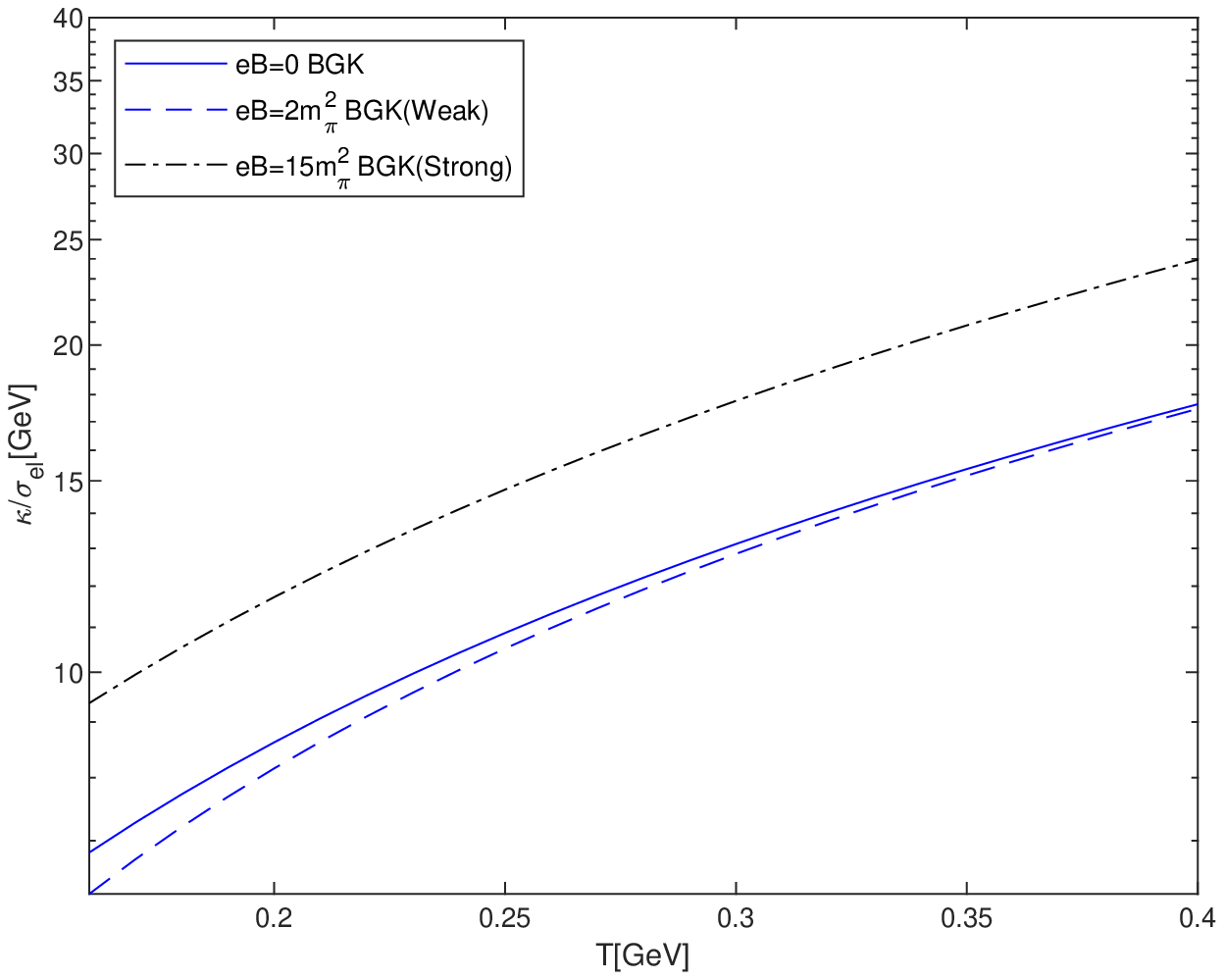}
		\caption{}\label{a018}
	\end{subfigure}
	\caption{The variation of the ratio of the thermal conductivity to the electrical conductivity ($\kappa/\sigma_{el}$) with temperature, (a) comparison with RTA model result (weak magnetic field) and (b) comparison with BGK model result (strong magnetic field).}\label{a009}
\end{figure}

\begin{figure}[ht]
	\begin{subfigure}{0.5\linewidth}
		\includegraphics[width=1\textwidth]{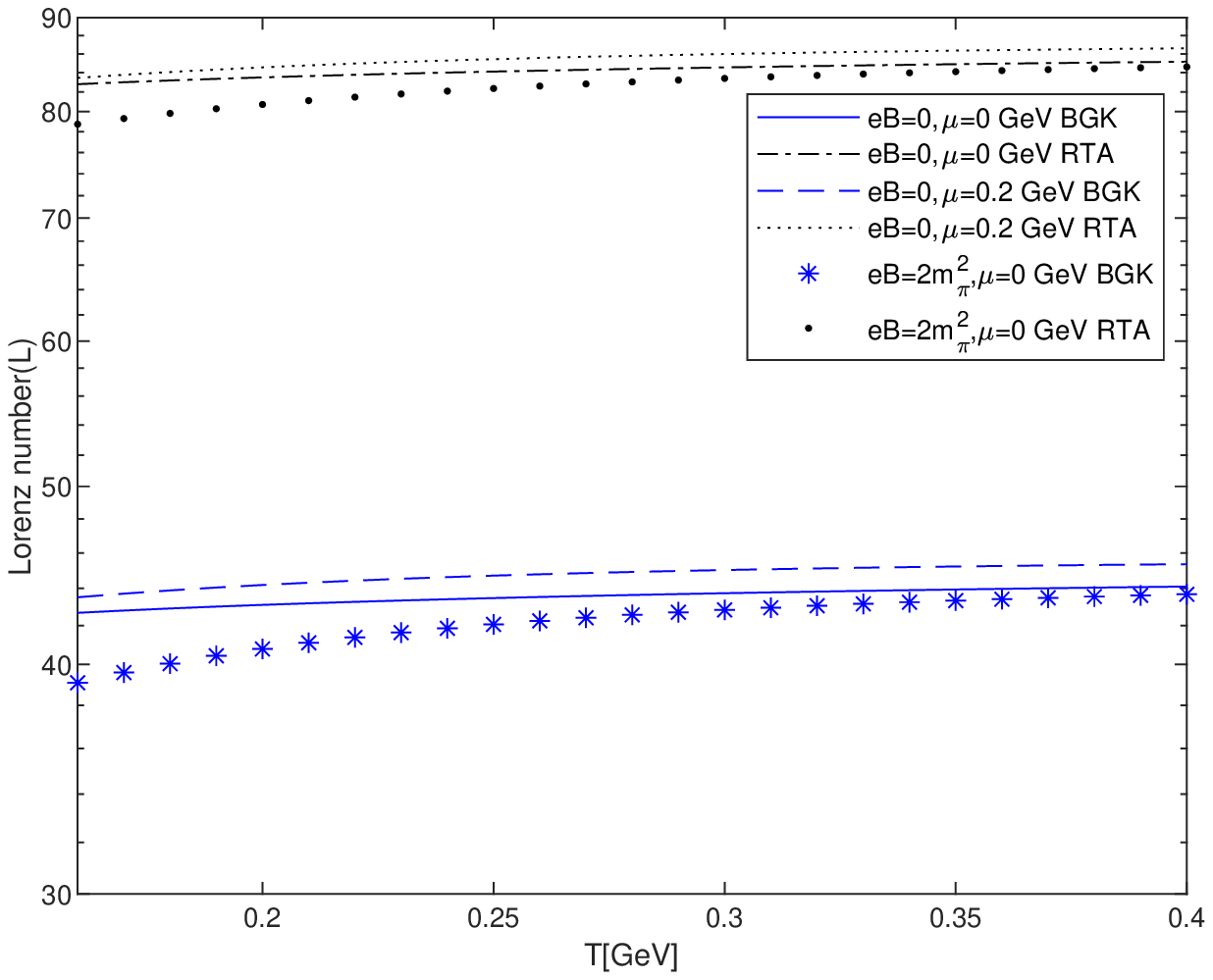}
		\caption{}\label{a019}
	\end{subfigure}
	\hfill
	\begin{subfigure}{0.5\linewidth}
		\includegraphics[width=1\textwidth]{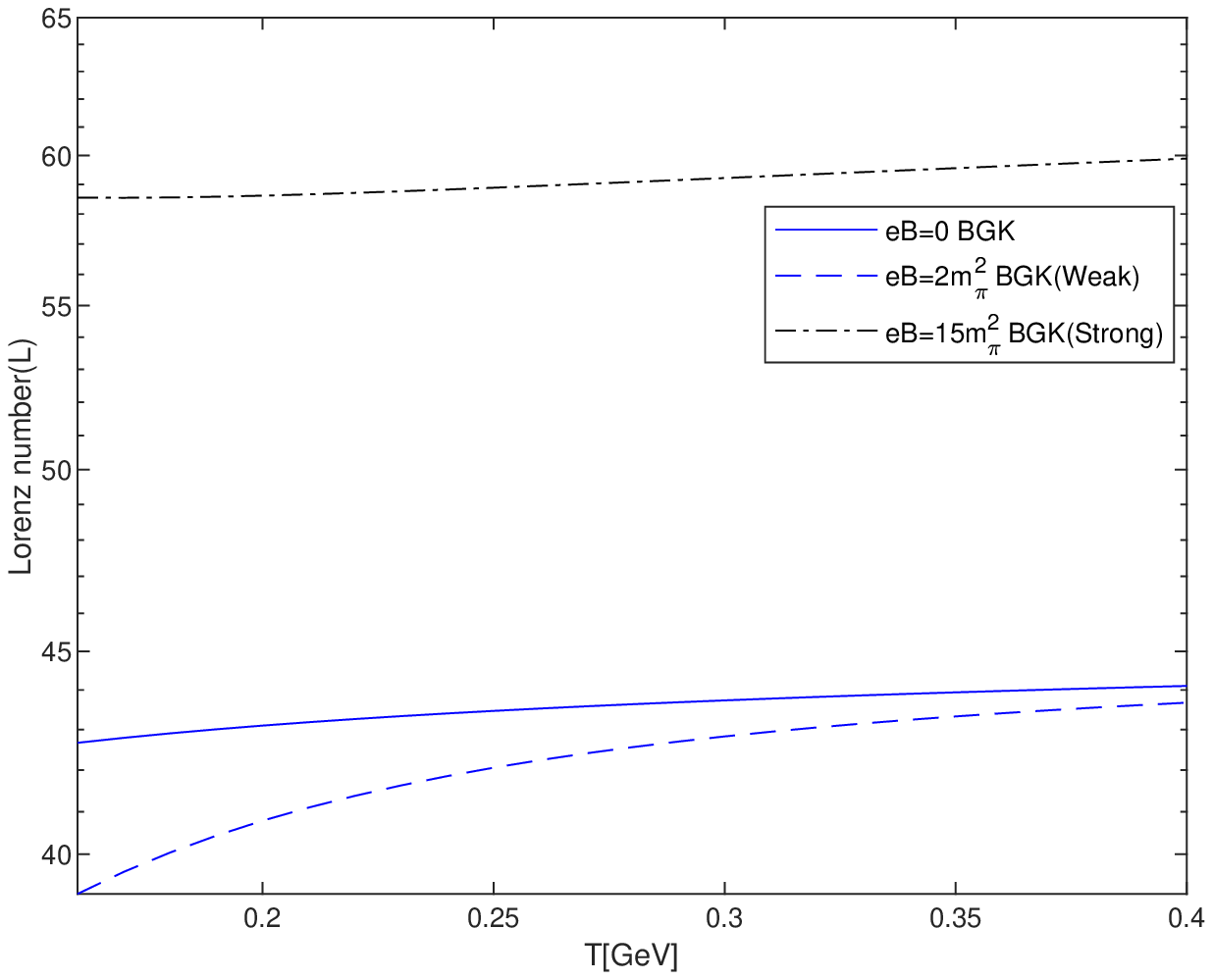}
		\caption{}\label{a020}
	\end{subfigure}
	\caption{The variation of the Lorenz number ($L$) with temperature, (a) comparison with RTA model result (weak magnetic field) and (b) comparison with BGK model result (strong magnetic field).}\label{a0010}
\end{figure}

This law indicates the relative behaviour of the charge and heat transport phenomena in a medium. In figures \ref{a017} and \ref{a018}, we show the variation of $\kappa/\sigma_{el}$ with temperature for different magnetic field strengths and chemical potentials and also compare this result with RTA (weak magnetic field) and BGK (strong magnetic field) model results, respectively. The same comparison has been done in figure \ref{a0010} (\ref{a019} and \ref{a020}) for the Lorenz number. It is found that the ratio $\kappa/\sigma_{el}$ varies almost linearly with temperature (figure \ref{a009}), indicating that the heat transport forges ahead of the charge transport as the temperature of the medium increases. For a better understanding of the relative behaviour of the aforesaid transport phenomena, the Lorenz number is seen to vary in a monotonically increasing manner for low  temperature, indicating the violation of the Wiedemann-Franz law. On the other hand, the Lorenz number saturates at higher temperature (figure \ref{a0010}). Figures \ref{a017} and \ref{a019} depict the domination of the relaxation collision integral over the BGK collision integral, whereas figures \ref{a018} and \ref{a020} indicate that the Lorenz number in the strong magnetic field regime prevails over the one in the weak magnetic field regime. 

\subsection{Knudsen number and specific heat}
Knudsen number ($\Omega$) is a dimensionless number defined as the ratio of the mean free path ($\lambda$) and the characteristic length scale of the medium ($l$), {\em i.e.}, 
\begin{equation}\label{O}
	\Omega=\frac{\lambda}{l}.
\end{equation}
If $\Omega$ is less than one ({\em i.e.} $\lambda<l$), then we can apply the equilibrium hydrodynamics to this system. So with the help of the Knudsen number, we can predict the local equilibrium of a system. The mean free path $\lambda$ in eq. $(\ref{O})$ is computed as
\begin{equation}
	\lambda=\frac{3\kappa}{VC_v},
\end{equation}
where $C_v$ and $V$ represent the specific heat at constant volume and relative speed, respectively. Specific heat is calculated with the help of the equation given below. 
\begin{equation}
	C_v=\frac{\partial \epsilon}{\partial T}.
\end{equation}
Here, $\epsilon$ is given in eq. ($\ref{P}$), and we have fixed the value of $V\approx1$ and $l=4$ fm. 

\begin{figure}[H]
	\begin{subfigure}{0.5\linewidth}
		\includegraphics[width=1\textwidth]{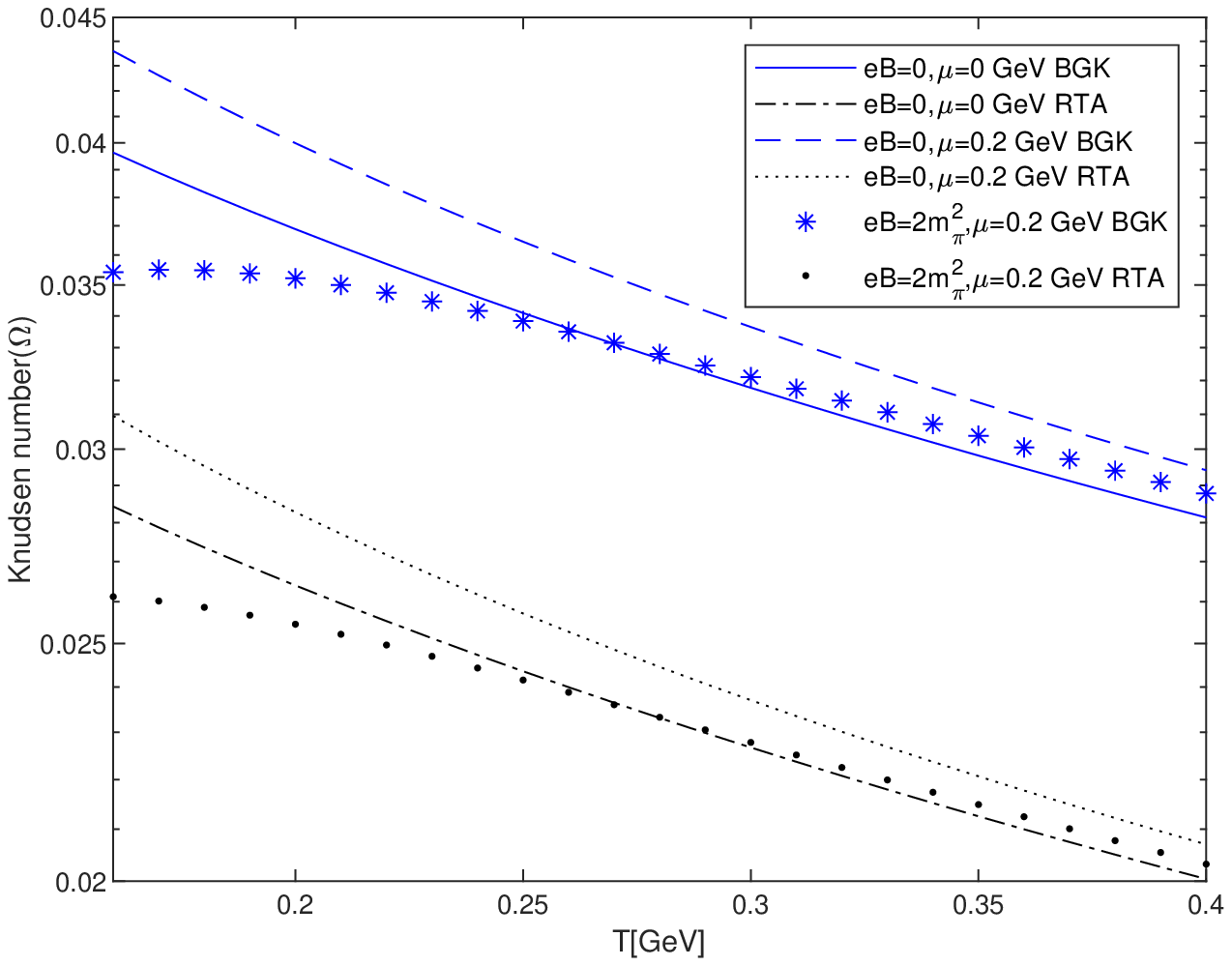}
		\caption{}\label{a021}
	\end{subfigure}
	\hfill
	\begin{subfigure}{0.5\linewidth}
		\includegraphics[width=1\textwidth]{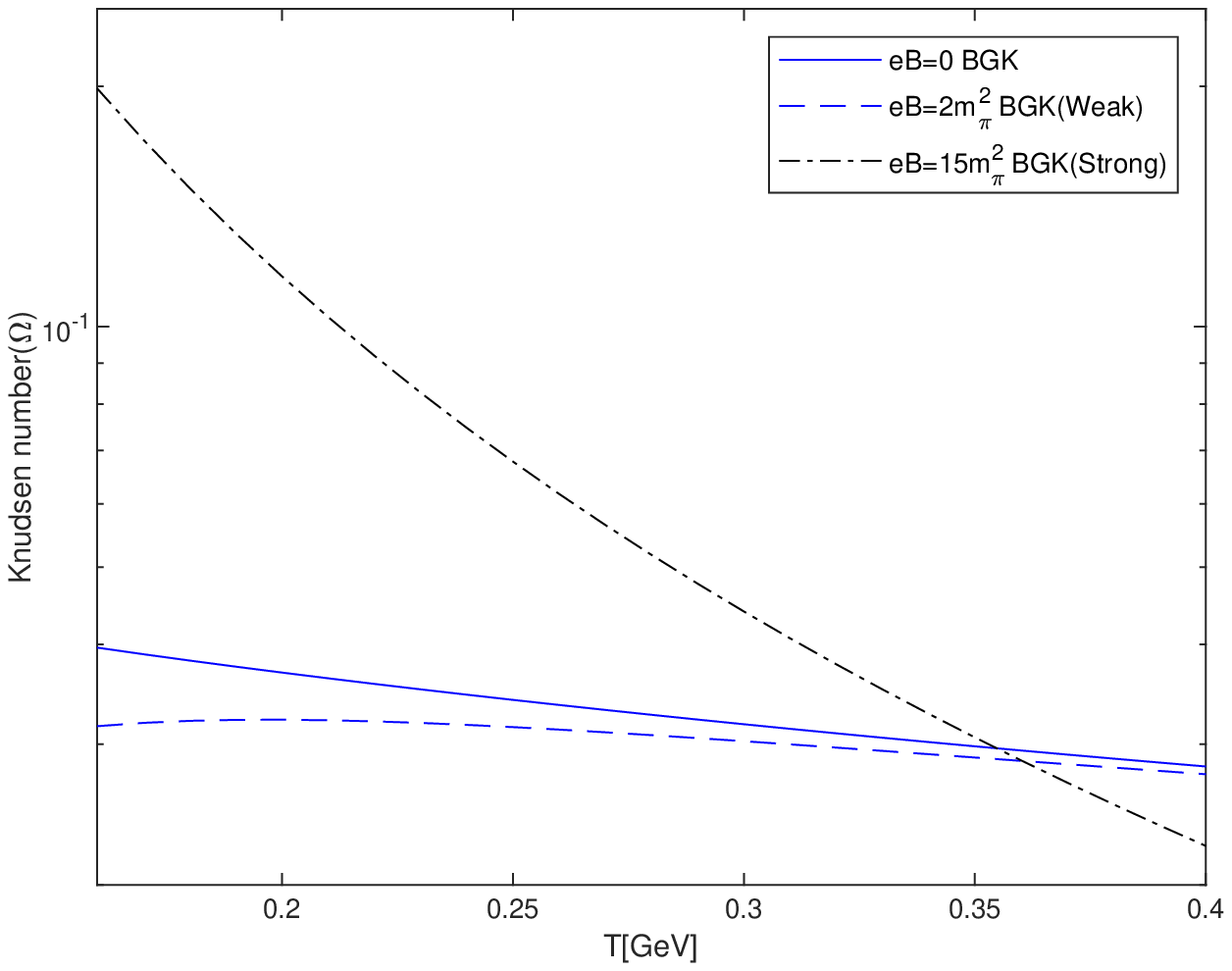}
		\caption{}\label{a022}
	\end{subfigure}
	\caption{The variation of the Knudsen number ($\Omega$) with temperature, (a) comparison with RTA model result (weak magnetic field) and (b) comparison with BGK model result (strong magnetic field).}\label{a0011}
\end{figure}

\begin{figure}[H]
	\begin{subfigure}{0.5\linewidth}
		\includegraphics[width=1\textwidth]{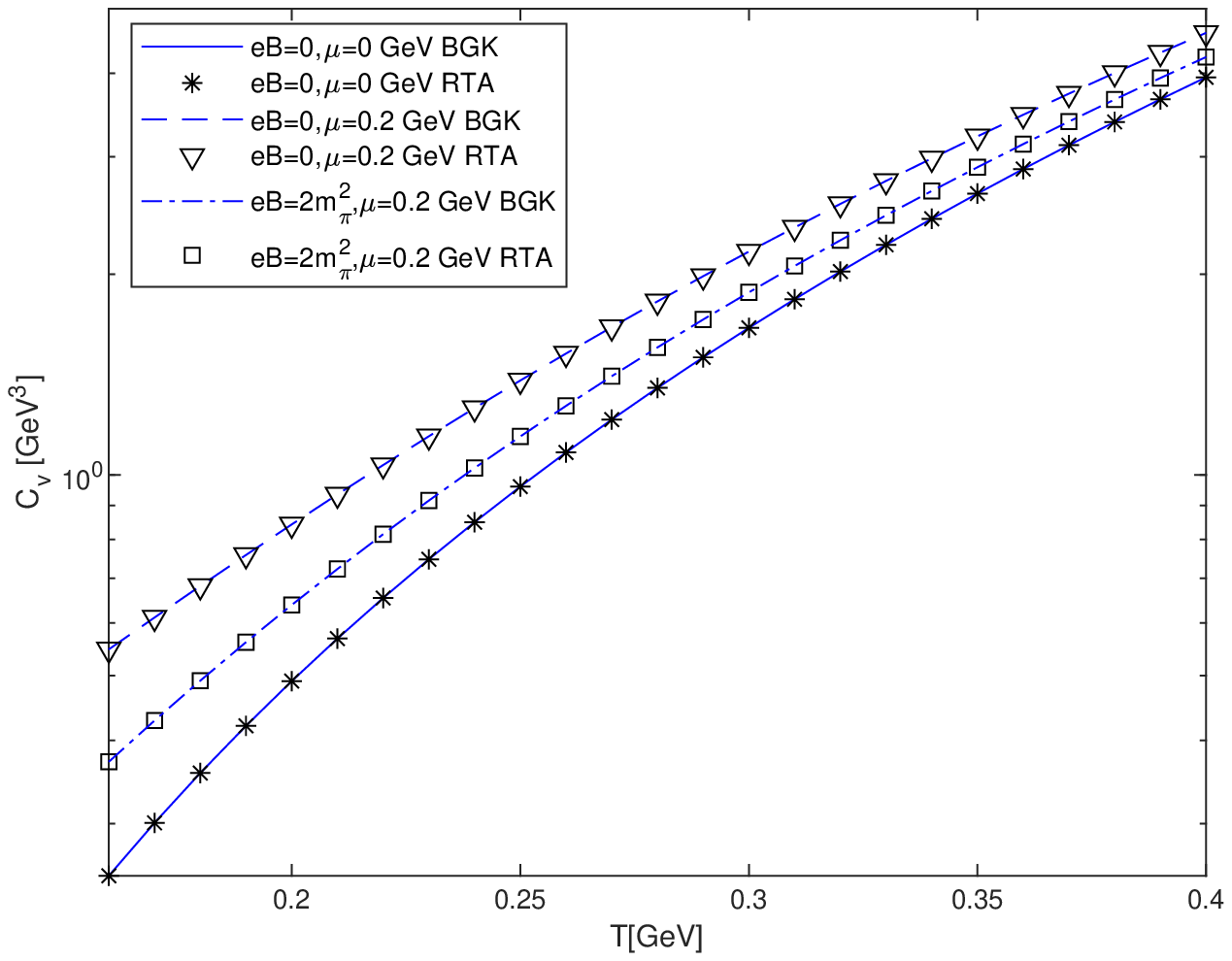}
		\caption{}\label{a023}
	\end{subfigure}
	\hfill
	\begin{subfigure}{0.5\linewidth}
		\includegraphics[width=1\textwidth]{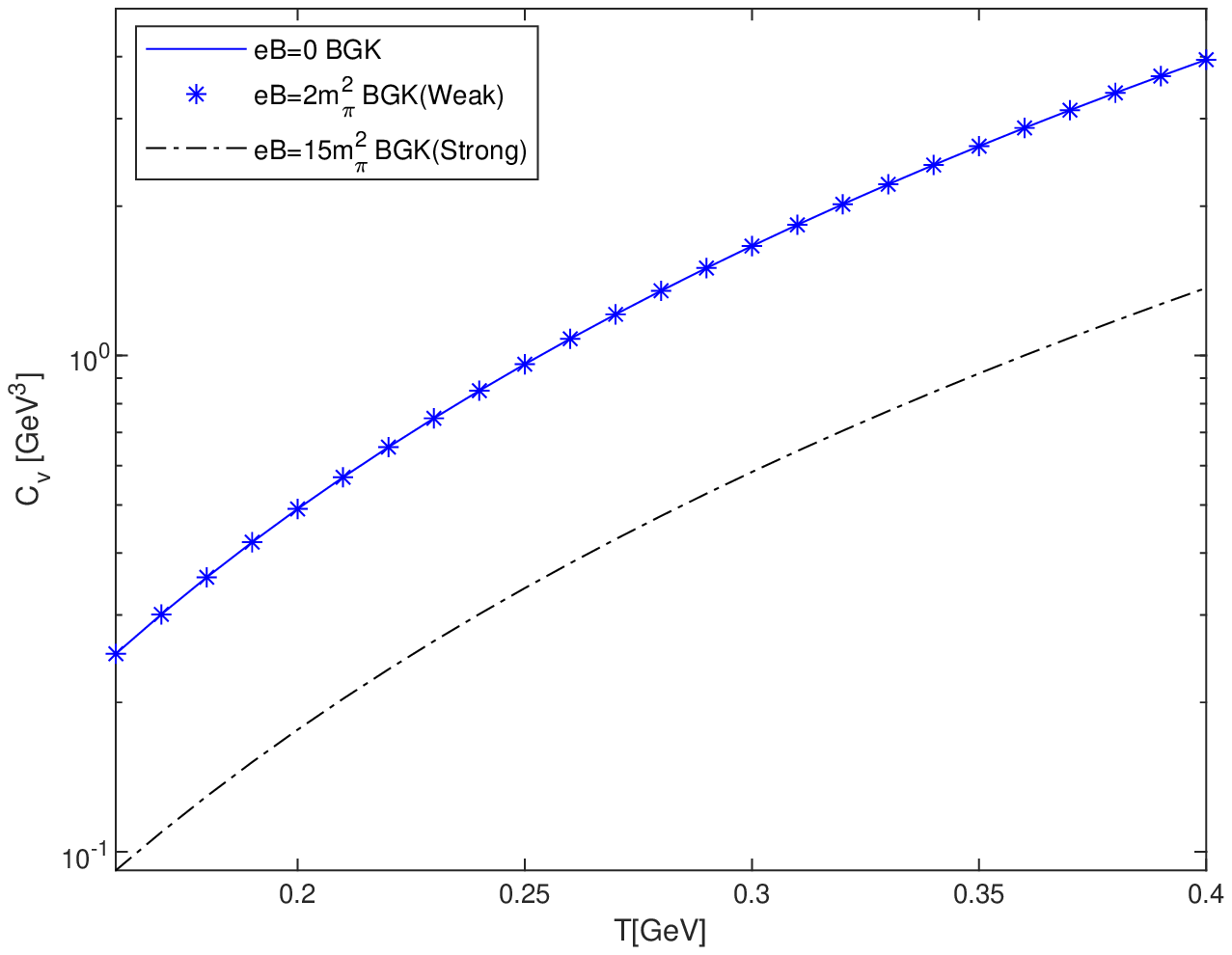}
		\caption{}\label{a024}
	\end{subfigure}
	\caption{The variation of the specific heat ($C_v$) with temperature, (a) comparison with RTA model result (weak magnetic field) and (b) comparison with BGK model result (strong magnetic field).}\label{a0012}
\end{figure}

Figures \ref{a0011} (\ref{a021} and \ref{a022}) and \ref{a0012} (\ref{a023} and \ref{a024}) show the variations of the Knudsen number and the specific heat with the temperature for different combinations of magnetic field and chemical potential, respectively as well as comparison with RTA (weak magnetic field) and BGK (strong magnetic field) model results. Overall, the Knudsen number shows a decreasing trend, whereas specific heat shows an increasing nature with temperature. From the above comparison, we can see that the modified BGK collision term is found to dominate over the RTA collision term, and the BGK model result in the strong magnetic field comes out to be larger than the corresponding weak magnetic field result, which can be understood from the behaviours of $\kappa$ (figure \ref{a007}) and $C_v$ (figure \ref{a0012}) in the similar environment. The value of $\Omega$, as seen in figure \ref{a0011} for all cases, is less than unity, which indicates the validity of the system being in local equilibrium even in the presence of both magnetic field and chemical potential. 

\subsection{Elliptic flow}
\begin{figure}[H]
	\begin{subfigure}{0.5\linewidth}
		\includegraphics[width=1\textwidth]{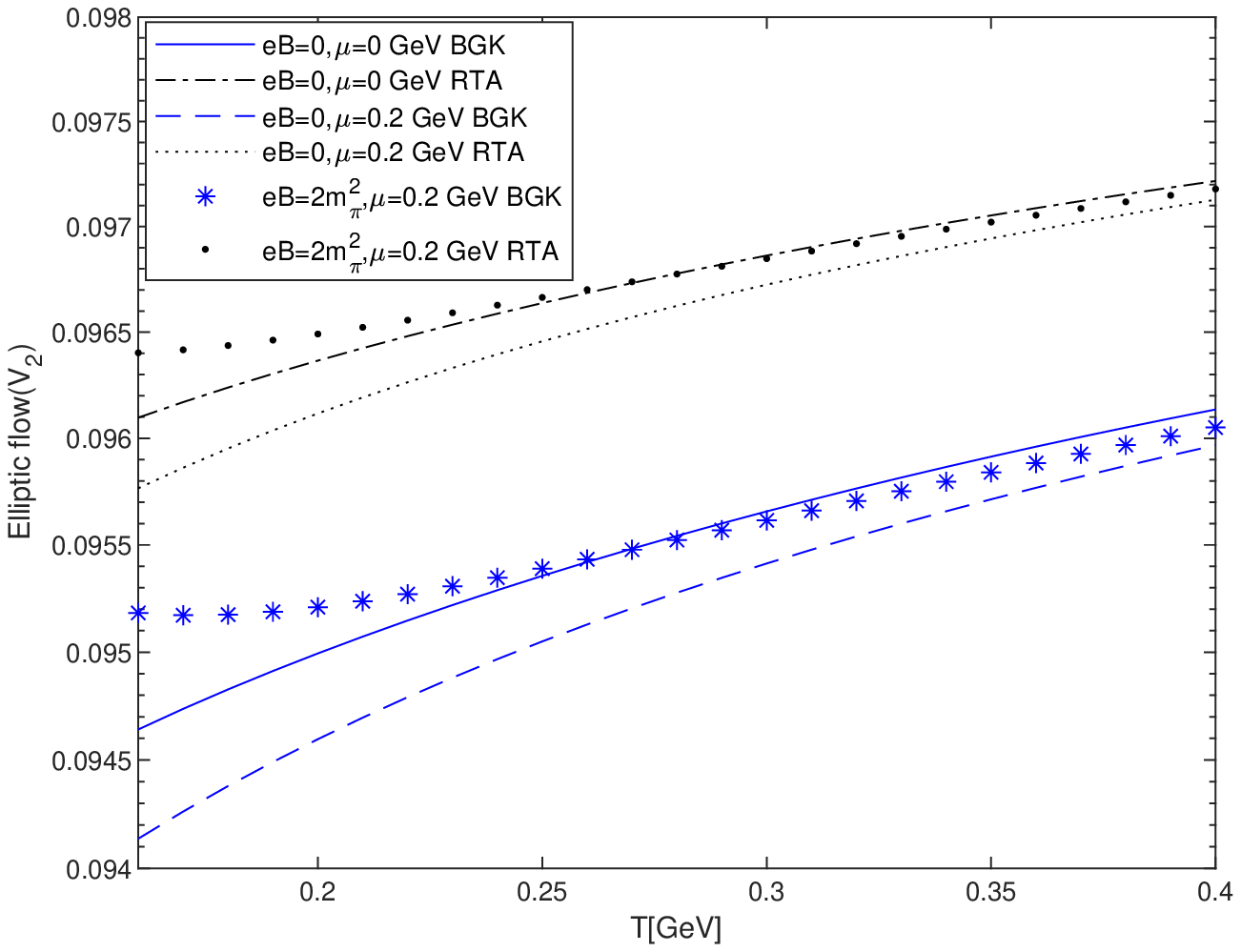}
		\caption{}\label{a025}
	\end{subfigure}
	\hfill
	\begin{subfigure}{0.5\linewidth}
		\includegraphics[width=1\textwidth]{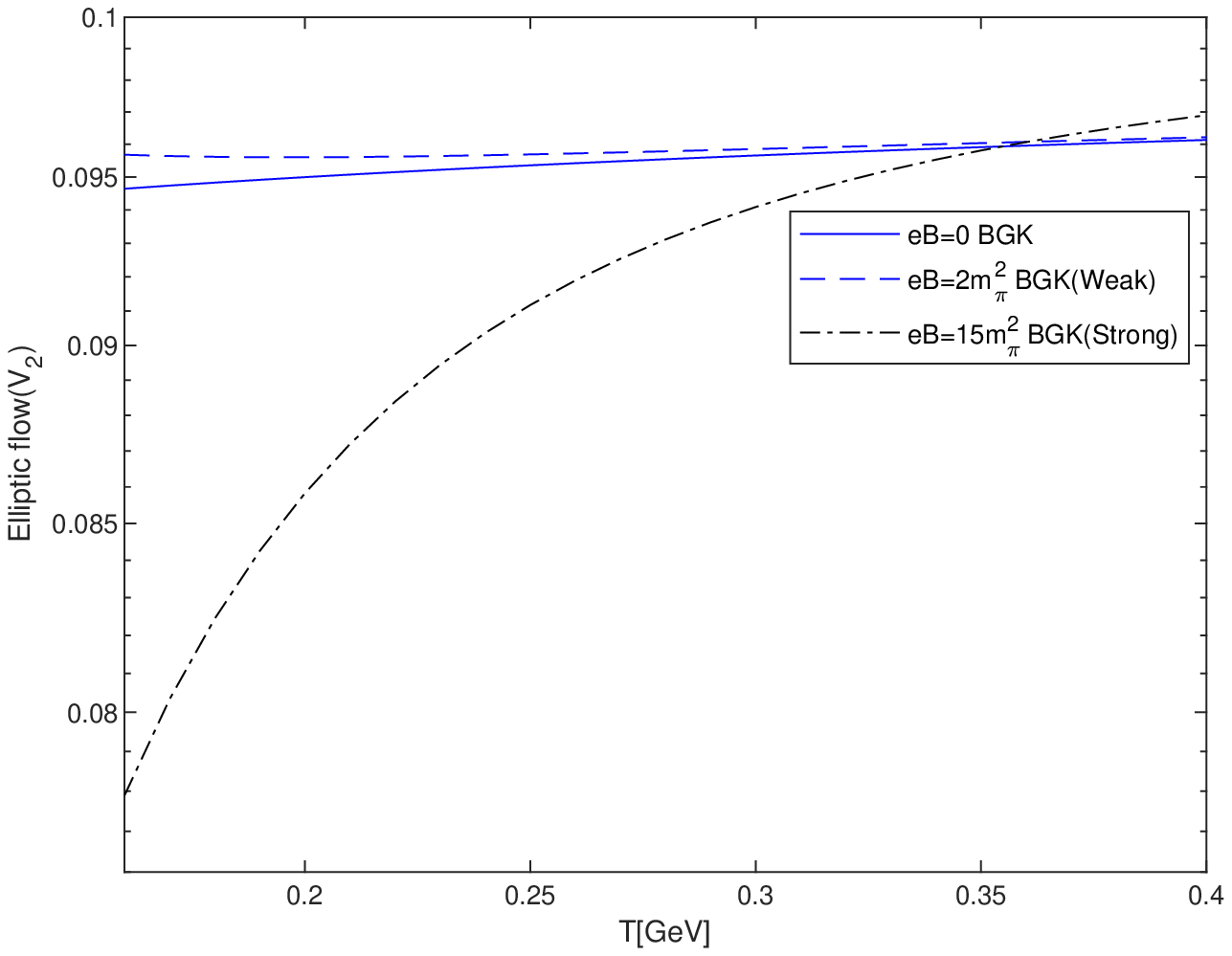}
		\caption{}\label{a026}
	\end{subfigure}
	\caption{The variation of the elliptic  flow coefficient	($v_2$) with temperature, (a) comparison with RTA model result (weak magnetic field) and (b) comparison with BGK model result (strong magnetic field).}\label{a0013}
\end{figure}

The elliptic flow coefficient ($v_2$),  describes the azimuthal anisotropy in the momentum space of produced particles in heavy ion collisions. The elliptic flow is related to the Knudsen number by the following equation \cite{bhalerao2005elliptic,drescher2007centrality,gombeaud2007elliptic}  as
\begin{equation}
	v_2=\frac{v^h_2}{1+\frac{\Omega}{\Omega_h}},
\end{equation}
where $v^h_2$ represents the value of elliptic flow coefficient in the hydrodynamic limit ({\em i.e.} $\Omega\to 0$ limit). The value of ${\Omega_h}$ can be calculated by observing the transition between the hydrodynamic regime and the free streaming particle regime. In our work, we have used $v^h_2\approx0.1$ and ${\Omega_h=0.7}$. Figure \ref{a0013} (\ref{a025} and \ref{a026}) shows the variation of elliptic flow coefficient with the temperature at different combinations of magnetic field and chemical potential as well as the comparison with RTA (weak magnetic field) and BGK (strong magnetic field) model results. It can be observed that $v_2$ shows an increasing trend with temperature for different combinations of weak magnetic field and finite chemical potential. From the above comparison, it is inferred that the RTA collision term dominates over the modified BGK collision term (figure \ref{a025}), and the value of the elliptic flow in BGK model for the strong magnetic field is less than its counterpart in the weak magnetic field (figure \ref{a026}), which can be understood from the behaviour of the Knudsen number with temperature as shown in figure \ref{a0011}. 

\section{Summary}
The effects of weak magnetic field and finite chemical potential on a strongly interacting thermal QCD medium  have been investigated through different charge and heat transport coefficients. Due to the complicated nature of the collision integral in the Boltzmann transport equation, we consider the problem in a kinetic theory approach with a modified collision integral under the BGK model. The magnetic field is assumed to be weak in comparison to the temperature scale in the system. In the presence of a magnetic field, the transport coefficients lose their isotropic nature and possess different components. Charge and heat transport coefficients such as, electrical conductivity ($\sigma_{el}$), Hall conductivity ($\sigma_{H}$), thermal conductivity ($\kappa$) and Hall-type thermal conductivity ($\kappa_{H}$) have been calculated by solving the relativistic Boltzmann transport equation. We have examined how the modified collision integral (BGK model) affects these transport coefficients in comparison to the often used relaxation type collision integral (RTA model). We also observed how the magnitudes of these coefficients change when we apply the quasiparticle model to the system, where the usual rest masses are being replaced by the medium generated masses. The modified BGK collision integral enhances both the charge transport ($\sigma_{el}$, $\sigma_{H}$) and the heat transport ($\kappa$, $\kappa_{H}$) coefficients as compared to the RTA collision integral in weak magnetic field. At finite chemical potential, all the charge and heat transport coefficients get enhanced, whereas, with the weak magnetic field, $\sigma_{el}$ and $\kappa$ decrease, and $\sigma_{H}$ and $\kappa_{H}$ increase. The transport coefficients were further used to study the Wiedemann-Franz law, the Lorenz number, the Knudsen number, the specific heat and the elliptic flow. The RTA collision integral dominates over the BGK collision term for the ratio of thermal-to-electrical conductivity (Wiedemann-Franz law). The Lorenz number is seen to increase monotonically with temperature, which indicates the violation of the Wiedemann-Franz law for a thermalised QCD medium in the presence of weak magnetic field. The magnitude of the Knudsen number remains below unity for both BGK and RTA models, which indicates the existence of a local equilibrium for the medium. The elliptic flow coefficient gets increased in the presence of weak magnetic field, whereas the presence of finite chemical potential decreases it. In addition, the elliptic flow coefficient in BGK model is observed to be less than that in RTA model in the presence of a weak magnetic field. 

\section{Acknowledgment}
One of us (S. R.) acknowledges the Indian Institute of Technology Bombay for the Institute 
postdoctoral fellowship. 

\appendix
\appendixpage
\addappheadtotoc
\begin{appendix}
\renewcommand{\theequation}{A.\arabic{equation}}
\section{Derivation of equation \eqref{x2}}\label{appendix A}
From equation (\ref{F}), we have
 \begin{align}\label{ansatz}
 	\begin{split}
	f_f=& n_f n^{-1}_{eq,f}f_{eq,f}-\tau_fq\textbf{E}.\frac{\partial f_{eq,f}}{\partial \textbf{p}}-\boldsymbol{\Gamma}. \frac{\partial f_{eq,f}}{\partial \textbf{p}},
	\\
	=& n_f n^{-1}_{eq,f}f_{eq,f}-\tau_fq E\frac{\partial f_{eq,f}}{\partial p_x}-{\Gamma_x}\frac{\partial f_{eq,f}}{\partial {p_x}}-{\Gamma_y} \frac{\partial f_{eq,f}}{\partial {p_y}}-{\Gamma_z} \frac{\partial f_{eq,f}}{\partial {p_z}}.
\end{split}
\end{align}
Now, after putting the values of $\frac{\partial f_{eq,f}}{\partial p_x}$, $\frac{\partial f_{eq,f}}{\partial {p_y}}$, $\frac{\partial f_{eq,f}}{\partial {p_z}}$ (as given by eq. (\ref{G})), ${\Gamma_x}$, ${\Gamma_y}$ and $\Gamma_z$ (as given by eq. (\ref{x3})) in eq. \eqref{ansatz}, we have
\begin{multline}
		f_f=n_f n^{-1}_{eq,f}f_{eq,f}-\tau_fq E\Biggl\lbrace -\beta v_x f_{eq,f}(1-f_{eq,f})\Biggr\rbrace -\Biggl\lbrace \frac{qE\tau_f(1-\omega^2_c \tau^2_f)}{(1+\omega^2_c \tau^2_f)}\Biggr\rbrace \\ \times\Biggl\lbrace -\beta v_x f_{eq,f}(1-f_{eq,f}) \Biggr\rbrace  -\Biggl\lbrace- \frac{2qE\omega_c \tau^2_f}{(1+\omega^2_c \tau^2_f)} \Biggr\rbrace \Biggr\lbrace -\beta v_yf_{eq,f} (1-f_{eq,f}) \Biggr\rbrace, \\
		f_f- n_f n^{-1}_{eq,f}f_{eq,f}=2qE\beta v_x\left( \frac{\tau_f}{(1+\omega^2_c \tau^2_f)}\right) f_{eq,f}(1-f_{eq,f})
		- 2qE\beta v_y \\ \times\left (\frac{\omega_c \tau^2_f}{(1+\omega^2_c \tau^2_f)}\right)f_{eq,f}(1-f_{eq,f}).
\end{multline}
For a proper treatment of the left-hand side of the above equation, we are considering an infinitesimal perturbation to the equilibrium distribution function: 
 $f_{eq,f}\rightarrow f_{eq,f}+\delta f_f$, $\delta f_f\ll f_{eq,f}$ and after linearizing it, we get
 \begin{multline}\label{x15}
 \delta f_f-g_f n^{-1}_{eq,f} f_{eq,f}\int_{p} \delta f_f=2qE\beta v_x\left( \frac{\tau_f}{(1+\omega^2_c \tau^2_f)}\right) f_{eq,f}(1-f_{eq,f})
 - 2qE\beta v_y \\ \times\left (\frac{\omega_c \tau^2_f}{(1+\omega^2_c \tau^2_f)}\right)f_{eq,f}(1-f_{eq,f}).
\end{multline}
Solving the above equation for $\delta f_{f}$ (neglecting the higher order, $\mathcal{O}((\delta f_{f})^2)$), we get
\begin{equation}\label{x4}
		\delta f_f=\delta f^{(0)}_f +g_f n^{-1}_{eq,f} f_{eq,f}\int_{p\prime} \delta f^{(0)}_f,
\end{equation}
where
\begin{multline}
	\delta f^{(0)}_f=2qE\beta v_x\left(\frac{\tau_f}{(1+\omega^2_c \tau^2_f)}\right) f_{eq,f}(1-f_{eq,f})
	- 2qE\beta v_y\left (\frac{\omega_c \tau^2_f}{(1+\omega^2_c \tau^2_f)}\right) \\ \times f_{eq,f}(1-f_{eq,f}).
\end{multline}
Putting it in eq. (\ref{x4}), we have
\begin{multline}\label{x5}	
	\delta f_f=\Biggl[ 2qE\beta v_x\Biggl( \frac{\tau_f}{(1+\omega^2_c \tau^2_f)}\Biggr) f_{eq,f}(1-f_{eq,f})
	- 2qE\beta v_y\left (\frac{\omega_c \tau^2_f}{(1+\omega^2_c \tau^2_f)}\right) \\ \times f_{eq,f}(1-f_{eq,f}) \Biggr]
	+ g_f n^{-1}_{eq,f} f_{eq,f} \int_{p\prime } \Biggr[ 2qE\beta v_x\Biggr( \frac{\tau_f}{(1+\omega^2_c \tau^2_f)}\Biggl) f_{eq,f}(1-f_{eq,f})\\
	- 2qE\beta v_y \Biggr (\frac{\omega_c \tau^2_f}{(1+\omega^2_c \tau^2_f)}\Biggl) f_{eq,f}(1-f_{eq,f}) \Biggl]. 
\end{multline}
	Similarly, for antiquarks, we have
\begin{multline}\label{x6}
	\delta {\bar{f}}_f=\Biggl[ 2\bar{q}E\beta v_x\Biggl( \frac{ \tau_{\bar{f}}}{(1+\omega^2_c \tau^2_{\bar{f}})}\Biggr) {\bar{f}}_{eq,f}(1-{\bar{f}}_{eq,f})
	- 2\bar{q}E\beta v_y\Biggl (\frac{\omega_c\tau^2_{\bar{f}}}{(1+\omega^2_c \tau^2_{\bar{f}})}\Biggr) \\ \times{\bar{f}}_{eq,f}(1-{\bar{f}}_{eq,f}) \Biggr]
	+ g_f n^{-1}_{eq,f} f_{eq,f} \int_{p \prime} \Biggl[ 2\bar{q}E\beta v_x\Biggl( \frac{\tau_{\bar{f}}}{(1+\omega^2_c \tau^2_{\bar{f}})}\Biggr) {\bar{f}}_{eq,f}(1-{\bar{f}}_{eq,f})\\
	- 2\bar{q}E\beta v_y\Biggl (\frac{\omega_c\tau^2_{\bar{f}}}{(1+\omega^2_c \tau^2_{\bar{f}})}\Biggr) {\bar{f}}_{eq,f}(1-{\bar{f}}_{eq,f}) \Biggr].
\end{multline}
After substituting the results of equations (\ref{x5}) and (\ref{x6}) in eq. (\ref{A}), we get
\begin{multline}\label{x7}
	J^{i}=\sum_{f}  g_{f}q_{f}\int \frac {d^3p}{(2\pi)^3} \frac{p_{\mu}}{\omega_f}\left[ (A+\bar A)+g_f n^{-1}_{eq,f} f_{eq,f}\int_{p\prime}(A+\bar A)\right]-\\\sum_{f}  g_{f}q_{f}\int \frac {d^3p}{(2\pi)^3} \frac{p_{\mu}}{\omega_f}\left[ (B+\bar B)+g_f n^{-1}_{eq,f} f_{eq,f}\int_{p\prime}(B+\bar B)\right],
\end{multline}
	where
	\begin{equation}
		A=2qE\beta v_x\Biggl( \frac{\tau_f}{(1+\omega^2_c \tau^2_f)}\Biggr) f_{eq,f}(1-f_{eq,f}),
	\end{equation}
			\begin{equation}
		B=2qE\beta v_y\left (\frac{\omega_c \tau^2_f}{(1+\omega^2_c \tau^2_f)}\right) \\ f_{eq,f}(1-f_{eq,f}),
	\end{equation}
	\begin{equation}
	\bar A=2\bar{q}E\beta v_x\Biggl( \frac{\tau_{\bar{f}}}{(1+\omega^2_c \tau^2_{\bar{f}})}\Biggr) {\bar{f}}_{eq,f}(1-{\bar{f}}_{eq,f}),
\end{equation}
	\begin{equation}
	\bar B=2\bar{q}E\beta v_y\Biggl (\frac{\omega_c\tau^2_{\bar{f}}}{(1+\omega^2_c \tau^2_{\bar{f}})}\Biggr) \\ {\bar{f}}_{eq,f}(1-{\bar{f}}_{eq,f}).
\end{equation}
Now recalling the eq. (\ref{C}), 
	\begin{equation}
	J^i=\sigma^{ij}E_j=(\sigma_{el}\delta^{ij}+\sigma_H\epsilon^{ij})E_j
\end{equation}
and comparing eq. (\ref{x7}) with this equation, we get the final expressions of electrical conductivity ($\sigma_{el}$) and Hall conductivity ($\sigma_{H}$) as given in equations (\ref{x8}) and (\ref{x9}), respectively. 

\renewcommand{\theequation}{B.\arabic{equation}}
\section{Derivation of equation \eqref{x10}}\label{appendix B}
After substituting the values of $\frac{\partial f_{eq,f}}{\partial p_x}$, $\frac{\partial f_{eq,f}}{\partial {p_y}}$, $\frac{\partial f_{eq,f}}{\partial {p_z}}$ (as given by eq. (\ref{G})), ${\Gamma_x}$, $\Gamma_y$ and ${\Gamma_z}$ (as given by equations (\ref{x12}), (\ref{x13}) and (\ref{x14}), respectively) in eq. \eqref{ansatz}, we get
\begin{multline}
	f_f=n_f n^{-1}_{eq,f}f_{eq,f}-\tau_fq E\Biggl\lbrace -\beta v_x f_{eq,f}(1-f_{eq,f})\Biggr\rbrace -\Biggl\lbrace qE\tau_{f}\frac{(1-\omega^2_c \tau^2_{f})}{(1+\omega^2_c \tau^2_{f})} \\ -\frac{\tau_{f}}{(1+\omega^2_c \tau^2_{f})}\frac{(\omega_f-h_f)}{T}\left( \partial^xT-\frac{T}{nh_f}\partial^xP\right)
	-\frac{\omega_c\tau^2_{f}}{(1+\omega^2_c \tau^2_{f})}\frac{(\omega_f-h_f)}{T} \left( \partial^yT-\frac{T}{nh_f}\partial^yP\right)\Biggr\rbrace \\ \times\Biggl\lbrace -\beta v_x f_{eq,f}(1-f_{eq,f}) \Biggr\rbrace  -\Biggl\lbrace -\frac{2qE\omega_c\tau^2_{f}}{(1+\omega^2_c \tau^2_{f})}-\frac{\tau_{f}}{(1+\omega^2_c \tau^2_{f})}\frac{(\omega_f-h_f)}{T} \left( \partial^yT-\frac{T}{nh_f}\partial^yP\right)\\
	+\frac{\omega_c\tau^2_{f}}{(1+\omega^2_c \tau^2_{f})}\frac{(\omega_f-h_f)}{T}\left( \partial^xT-\frac{T}{nh_f}\partial^xP\right)\Biggr\rbrace \Biggr\lbrace -\beta v_yf_{eq,f} (1-f_{eq,f}) \Biggr\rbrace, \\ f_f- n_f n^{-1}_{eq,f}f_{eq,f}=\frac{2qE\tau_fv_x \beta}{(1+\omega^2_c \tau^2_{f})}f_{eq,f}(1-f_{eq,f})-\frac{2qE\omega_c\tau^2_f v_y \beta}{(1+\omega^2_c \tau^2_{f})}f_{eq,f}(1-f_{eq,f})\\
	-\beta f_{eq,f}(1-f_{eq,f})\frac{\tau_{f}}{(1+\omega^2_c \tau^2_{f})}\frac{(\omega_f-h_f)}{T}\left[ v_x\left( \partial^xT-\frac{T}{nh_f}\partial^xP\right)+v_y\left( \partial^yT-\frac{T}{nh_f}\partial^yP\right)\right]\\-\beta f_{eq,f}(1-f_{eq,f})\frac{\omega_c\tau^2_{f}}{(1+\omega^2_c \tau^2_{f})}\frac{(\omega_f-h_f)}{T}\left[ v_x\left( \partial^yT-\frac{T}{nh_f}\partial^yP\right)-v_y\left( \partial^xT-\frac{T}{nh_f}\partial^xP\right)\right].
\end{multline}
Following the previous steps (as done in equations (\ref{x15}) and (\ref{x4})), we have
	\begin{multline}
	\delta f^{(0)}_f=\frac{2qE\tau_fv_x \beta}{(1+\omega^2_c \tau^2_{f})}f_{eq,f}(1-f_{eq,f})-\frac{2qE\omega_c\tau^2_f v_y \beta}{(1+\omega^2_c \tau^2_{f})}f_{eq,f}(1-f_{eq,f})\\
	-\beta f_{eq,f}(1-f_{eq,f})\frac{\tau_{f}}{(1+\omega^2_c \tau^2_{f})}\frac{(\omega_f-h_f)}{T}\left[ v_x\left( \partial^xT-\frac{T}{nh_f}\partial^xP\right)+v_y\left( \partial^yT-\frac{T}{nh_f}\partial^yP\right)\right]\\-\beta f_{eq,f}(1-f_{eq,f})\frac{\omega_c\tau^2_{f}}{(1+\omega^2_c \tau^2_{f})}\frac{(\omega_f-h_f)}{T}\left[ v_x\left( \partial^yT-\frac{T}{nh_f}\partial^yP\right)-v_y\left( \partial^xT-\frac{T}{nh_f}\partial^xP\right)\right],
\end{multline}
 \begin{multline}\label{x16}
 	\delta f_f=\frac{2qE\tau_fv_x \beta}{(1+\omega^2_c \tau^2_{f})}f_{eq,f}(1-f_{eq,f})-\frac{2qE\omega_c\tau^2_f v_y \beta}{(1+\omega^2_c \tau^2_{f})}f_{eq,f}(1-f_{eq,f})\\
 	-\beta f_{eq,f}(1-f_{eq,f})\frac{\tau_{f}}{(1+\omega^2_c \tau^2_{f})}\frac{(\omega_f-h_f)}{T}\left[ v_x\left( \partial^xT-\frac{T}{nh_f}\partial^xP\right)+v_y\left( \partial^yT-\frac{T}{nh_f}\partial^yP\right)\right]\\-\beta f_{eq,f}(1-f_{eq,f})\frac{\omega_c\tau^2_{f}}{(1+\omega^2_c \tau^2_{f})}\frac{(\omega_f-h_f)}{T}\left[ v_x\left( \partial^yT-\frac{T}{nh_f}\partial^yP\right)-v_y\left( \partial^xT-\frac{T}{nh_f}\partial^xP\right)\right]\\+ g_f n^{-1}_{eq,f} f_{eq,f}\int_{ p\prime} \biggl[ \frac{2qE\tau_fv_x \beta}{(1+\omega^2_c \tau^2_{f})}f_{eq,f}(1-f_{eq,f})-\frac{2qE\omega_c\tau^2_f v_y \beta}{(1+\omega^2_c \tau^2_{f})}f_{eq,f}(1-f_{eq,f})\\
 	-\beta f_{eq,f}(1-f_{eq,f})\frac{\tau_{f}}{(1+\omega^2_c \tau^2_{f})}\frac{(\omega_f-h_f)}{T}\left[ v_x\left( \partial^xT-\frac{T}{nh_f}\partial^xP\right)+v_y\left( \partial^yT-\frac{T}{nh_f}\partial^yP\right)\right]\\-\beta f_{eq,f}(1-f_{eq,f})\frac{\omega_c\tau^2_{f}}{(1+\omega^2_c \tau^2_{f})}\frac{(\omega_f-h_f)}{T}\left[ v_x\left( \partial^yT-\frac{T}{nh_f}\partial^yP\right)-v_y\left( \partial^xT-\frac{T}{nh_f}\partial^xP\right)\right]\biggr] , 
 \end{multline}
and
\begin{multline}\label{x17}
	\delta \bar{f}_f=\frac{2qE\tau_{\bar{f}}v_x \beta}{(1+\omega^2_c \tau^2_{\bar{f}})}\bar{f}_{eq,f}(1-\bar{f}_{eq,f})-\frac{2qE\omega_c\tau^2_{\bar{f}} v_y \beta}{(1+\omega^2_c \tau^2_{\bar{f}})}\bar{f}_{eq,f}(1-\bar{f}_{eq,f})\\
	-\beta \bar{f}_{eq,f}(1-\bar{f}_{eq,f})\frac{\tau_{\bar{f}}}{(1+\omega^2_c \tau^2_{\bar{f}})}\frac{(\omega_f-h_{\bar{f}})}{T}\left[ v_x\left( \partial^xT-\frac{T}{nh_f}\partial^xP\right)+v_y\left( \partial^yT-\frac{T}{nh_f}\partial^yP\right)\right]\\-\beta \bar{f}_{eq,f}(1-\bar{f}_{eq,f})\frac{\omega_c\tau^2_{\bar{f}}}{(1+\omega^2_c \tau^2_{\bar{f}})}\frac{(\omega_f-h_{\bar{f}})}{T}\left[ v_x\left( \partial^yT-\frac{T}{nh_f}\partial^yP\right)-v_y\left( \partial^xT-\frac{T}{nh_f}\partial^xP\right)\right]\\+ g_f n^{-1}_{eq,f} \bar{f}_{eq,f}\int_{p \prime } \biggl[ \frac{2qE\tau_{\bar{f}}v_x \beta}{(1+\omega^2_c \tau^2_{\bar{f}})}\bar{f}_{eq,f}(1-\bar{f}_{eq,f})-\frac{2qE\omega_c\tau^2_{\bar{f}} v_y \beta}{(1+\omega^2_c \tau^2_{\bar{f}})}\bar{f}_{eq,f}(1-\bar{f}_{eq,f})\\
	-\beta \bar{f}_{eq,f}(1-\bar{f}_{eq,f})\frac{\tau_{\bar{f}}}{(1+\omega^2_c \tau^2_{\bar{f}})}\frac{(\omega_f-h_{\bar{f}})}{T}\left[ v_x\left( \partial^xT-\frac{T}{nh_f}\partial^xP\right)+v_y\left( \partial^yT-\frac{T}{nh_f}\partial^yP\right)\right]\\-\beta \bar{f}_{eq,f}(1-\bar{f}_{eq,f})\frac{\omega_c\tau^2_{\bar{f}}}{(1+\omega^2_c \tau^2_{\bar{f}})}\frac{(\omega_f-h_{\bar{f}})}{T}\left[ v_x\left( \partial^yT-\frac{T}{nh_f}\partial^yP\right)-v_y\left( \partial^xT-\frac{T}{nh_f}\partial^xP\right)\right]\biggr] .	
\end{multline}
After putting the results of equations (\ref{x16}) and (\ref{x17}) in eq. (\ref{K}), we get
\begin{multline}\label{x18}
		Q^{i}=\sum_{f} g_f \int \frac{d^3p}{(2\pi)^3}\frac{p^i}{\omega_f}\Biggl[ \Bigl(\omega_f -h_f\Bigr)\Bigl[\Bigl(2qEv_x\beta A-2qEv_y\beta B-\beta^2C-\beta^2D\Bigr)\\  + g_f n^{-1}_{eq,f} f_{eq,f}\int_{ p\prime}\Bigl( 2qEv_x\beta A-2qEv_y\beta B-\beta^2C-\beta^2D\Bigr)\Bigr] +\Bigl(\omega_f -\bar {h}_f\Bigr)\\ \times \Bigl[\Bigl( 2qEv_x\beta \bar{A}-2qEv_y\beta\bar{B}-\beta^2\bar{C}-\beta^2\bar{D}\Bigr)  +g_f n^{-1}_{eq,f} f_{eq,f} \\ \times\int_{ p\prime}\Bigl( 2qEv_x\beta \bar{A}-2qEv_y\beta \bar{B}-\beta^2\bar{C}-\beta^2\bar{D}\Bigr) \Bigr]\Biggr],
\end{multline}
where
\begin{equation}
	A=\frac{\tau_f}{(1+\omega^2_c \tau^2_{f})}f_{eq,f}(1-f_{eq,f}),
\end{equation}
\begin{equation}
	B=\frac{\omega_c\tau^2_f}{(1+\omega^2_c \tau^2_{f})}f_{eq,f}(1-f_{eq,f}),
\end{equation}
\begin{equation}
	C=\frac{\tau_{f}}{(1+\omega^2_c \tau^2_{f})}(\omega_f-h_f)f_{eq,f}(1-f_{eq,f})\left[ v_x\left( \partial^xT-\frac{T}{nh_f}\partial^xP\right)+v_y\left( \partial^yT-\frac{T}{nh_f}\partial^yP\right)\right],
\end{equation}
\begin{equation}
	D=\frac{\omega_c\tau^2_{f}}{(1+\omega^2_c \tau^2_{f})}(\omega_f-h_f)f_{eq,f}(1-f_{eq,f})\left[ v_x\left( \partial^yT-\frac{T}{nh_f}\partial^yP\right)-v_y\left( \partial^xT-\frac{T}{nh_f}\partial^xP\right)\right],
\end{equation}
\\
\begin{equation}
	\bar{A}=\frac{\tau_{\bar{f}}}{(1+\omega^2_c \tau^2_{\bar{f}})}\bar{f}_{eq,f}(1-\bar{f}_{eq,f}),
\end{equation}
\begin{equation}
	\bar{B}=\frac{\omega_c\tau^2_{\bar{f}}}{(1+\omega^2_c \tau^2_{\bar{f}})}\bar{f}_{eq,f}(1-\bar{f}_{eq,f}),
\end{equation}
\begin{equation}
	\bar{C}=\frac{\tau_{\bar{f}}}{(1+\omega^2_c \tau^2_{\bar{f}})}(\omega_f-h_{\bar{f}})\bar{f}_{eq,f}(1-\bar{f}_{eq,f})\left[ v_x\left( \partial^xT-\frac{T}{nh_f}\partial^xP\right)+v_y\left(\partial^yT-\frac{T}{nh_f}\partial^yP\right)\right],
\end{equation}
\begin{equation}
	\bar{D}=\frac{\omega_c\tau^2_{\bar{f}}}{(1+\omega^2_c \tau^2_{\bar{f}})}(\omega_f-h_{\bar{f}})\bar{f}_{eq,f}(1-\bar{f}_{eq,f})\left[ v_x\left( \partial^yT-\frac{T}{nh_f}\partial^yP\right)-v_y\left( \partial^xT-\frac{T}{nh_f}\partial^xP\right)\right].
\end{equation}
Now recalling eq. (\ref{J}), 
\begin{equation}
	Q^i=-(\kappa\delta^{ij}+\kappa_H\epsilon^{ij})\left[\partial_jT-\frac{T}{(n_{eq,f} h_f)}\partial_jP\right]
\end{equation}
and comparing eq. (\ref{x18}) with this equation, we get the final expressions of thermal conductivity ($\kappa$) and Hall-type thermal conductivity ($\kappa_{H}$) as given in equations (\ref{x19}) and 
(\ref{x20}), respectively. 

\end{appendix}

	\bibliographystyle{unsrt}
	\bibliography{myref0072}
		
\end{document}